\documentclass{emulateapj}

\usepackage{amsmath}

\newcommand{\be}{\begin{displaymath}}
\newcommand{\ee}{\end{displaymath}}

\def\lsim{\hbox{\rlap{\raise 0.425ex\hbox{$<$}}\lower 0.65ex\hbox{$\sim$}}}
\def\gsim{\hbox{\rlap{\raise 0.425ex\hbox{$>$}}\lower 0.65ex\hbox{$\sim$}}}

\def\arcdeg{\mbox{$^\circ$}}

\shorttitle{Three-Parameter Relationship in SN~Ia: $M_p$, $t_r$, $v$}
\shortauthors{Zheng et al.}

\begin{document}

\title{An Empirical Fitting Method to Type Ia Supernova Light Curves. III. A Three-Parameter Relationship: Peak Magnitude, Rise Time, and Photospheric Velocity}

\author{WeiKang Zheng\altaffilmark{1,2},
Patrick L. Kelly\altaffilmark{1,3}, and
Alexei V. Filippenko\altaffilmark{1,4}
}

\altaffiltext{1}{Department of Astronomy, University of California, Berkeley, CA 94720-3411, USA.}
\altaffiltext{2}{e-mail: zwk@astro.berkeley.edu .}
\altaffiltext{3}{School of Physics and Astronomy, University of Minnesota, 116 Church Street SE, Minneapolis, MN 55455, USA}
\altaffiltext{4}{Miller Senior Fellow, Miller Institute for Basic Research in Science, University of California, Berkeley, CA 94720, USA}

\begin{abstract}
We examine the relationship between three parameters of Type Ia supernovae (SNe~Ia):
peak magnitude, rise time, and photospheric velocity at the time of peak brightness.
The peak magnitude is corrected for extinction using an estimate determined from MLCS2k2 fitting.
The rise time is measured from the well-observed $B$-band light curve with the
first detection at least 1~mag fainter than the peak magnitude, and the photospheric velocity
is measured from the strong absorption feature of Si~II~$\lambda$6355 at the time of
peak brightness. 
We model the relationship among these three parameters using an expanding fireball with two assumptions:
(a) the optical emission is approximately that of a blackbody, and (b) the photospheric temperatures
of all SNe~Ia are the same at the time of peak brightness. We compare the precision of the
distance residuals inferred using this physically motivated model
against those from the empirical Phillips relation and 
the MLCS2k2 method for 47 low-redshift SNe~Ia ($0.005 < z< 0.04$)
and find comparable scatter.
However, SNe~Ia in our sample with higher velocities are inferred to be intrinsically fainter.
Eliminating the high-velocity SNe and applying a more stringent extinction cut
to obtain a ``low-$v$ golden sample'' of 22 SNe, we 
obtain significantly reduced scatter of $0.108 \pm 0.018$~mag in the new relation, better than those of the Phillips relation and the MLCS2k2 method.
For 250\,km\,s$^{-1}$ of residual peculiar motions, we find 68\% and 95\% upper limits on the intrinsic scatter of 0.07 and 0.10 mag, respectively.
\end{abstract}



\section{Introduction}\label{s:intro}

The correlation between the light-curve decline rate and the peak luminosity of Type Ia supernovae (SNe~Ia),
known as the ``Phillips relation'' (Phillips 1993; Phillips et al. 1999), allows SNe~Ia to be used
as standardizable candles with many important applications,
including measurements of the expansion history that reveal the acceleration of the Universe (Riess et al. 1998; Perlmutter et al. 1999). 
In the past decade, various efforts have improved the
distance estimation, with different parameters adopted to reduce the scatter --- for example, with
light curves through different filters (Riess et al. 1996;
 Wang et al. 2003, 2005; Tripp 1998; Guy et al. 2005, 2007; Jha
et al. 2007), information about the host galaxies (e.g., Kelly et al. 2010, 2015; Sullivan et al. 2010;
Lampeitl et al. 2010; Childress et al. 2013; Rigault et al. 2013, 2015),
spectroscopic features (e.g., Wang et al. 2009; Foley \& Kasen 2011; Blondin et al. 2011; 
Fakhouri et al. 2015),
and information about color (Wang et al. 2003; Conley et al. 2006) and color-stretch (Burns et al. 2014).
Typically, these methods are able to determine the luminosity of individual SNe~Ia with
an accuracy of 0.14--0.20~mag,
or potentially even as low as 0.07~mag (Wang et al. 2009; Foley \& Kasen 2011; 
Rigault et al. 2013, 2015; Kelly et al. 2015; Fakhouri et al. 2015).

In this paper, we introduce a new three-parameter relationship between the peak magnitude,
the rise time, and the photospheric velocity at the time of peak brightness, and we apply this
to a low-redshift SN~Ia sample. This relation may
indicate a new direction for further improving the cosmological utility of SNe~Ia.

\section{Motivation}\label{s:motivation}

Assuming that the SN~Ia luminosity scales with the surface area of the expanding fireball (which
is thought to be approximately a blackbody at early times), and given that optical wavelengths are on the
Rayleigh-Jeans tail of the nearly thermal spectrum, the SN~Ia luminosity
increases quadratically with photospheric radius (see Riess et al. 1999):
\begin{equation}
L \propto R^2 T \propto [v(t-t_0)]^2 T,
\label{eq_lvtt2}
\end{equation}
where $L$ is the SN luminosity, $R$ is the photospheric radius, $T$ is the fireball temperature,
$v$ is the photospheric expansion velocity, $t_0$ is the time of first light, and $t-t_0$ is the time
after first light. Zheng \& Filippenko (2017) have shown that, with some further assumptions and
replacing a constant photospheric velocity $v$ with a broken-power-law function,
the resulting function (Equation~7 of Zheng \& Filippenko 2017) can well fit
SN~Ia light curves. A good application of the fitting method is for estimating the first-light time
and the rise time of SNe~Ia, as shown by Zheng et al. (2017), where we use
the light curve measured about a week to 10 days after the first-light time 
for a sample of 56 SNe~Ia. 

Setting the time in Equation~\ref{eq_lvtt2} to $t_p$, the time at peak brightness\footnote{In principle, one could 
choose any specific time, but the time at peak brightness is the most convenient
and relates to measured parameters.}, one has
\begin{equation}
L_p \propto [(v_p) (t_p-t_0)]^2 T_p.
\label{eq_lvt2}
\end{equation}
In addition, one can define $t_p-t_0$ as
the rise time $t_{r}$ (namely, the time since first light until the 
time of peak brightness in different filters), giving
\begin{equation}
L_p \propto (v_p t_r)^2 T_p.
\label{eq_lvt2p}
\end{equation}

Assuming that the temperature at the time of peak brightness
is constant among different SNe~Ia, this becomes
\begin{equation}
L_p \propto (v_p t_r)^2,
\label{eq_lv2p}
\end{equation}
giving a clear relationship between the peak luminosity ($L_p$), the photospheric velocity at the time of peak brightness ($v_p$), and the rise time ($t_r$). 
Taking the logarithm of both sides and transforming units into the
magnitude system leads to
\begin{equation}
M_p = -2.5\,{\rm log}[(v_p t_r)^2] + C,
\label{eq_lvt2pm}
\end{equation}
where $M_p$ is the absolute magnitude at peak brightness and $C$ is a constant. Equation~\ref{eq_lvt2pm}
suggests that the peak magnitude of SNe~Ia is directly related to the photospheric velocity at peak brightness
($v_p$) and the rise time ($t_r$). This is the new relation we are presenting and
discussing here. For purposes of convenience, we assign the label
$M_{v^2t^2}$ to represent $-2.5\,{\rm log}[(v_p t_r)^2] + C$.

Note that in the above Equation~\ref{eq_lvtt2}, we have assumed optical wavelengths are on the
Rayleigh-Jeans tail of the nearly thermal spectrum, but the $B$ broadband filter, which is the band we
adopted for the following analysis, has a central wavelength that is too blue to fall on the Rayleigh-Jeans tail
for the temperatures of SNe~Ia near maximum light; thus, the absolute SN~Ia $B$-band magnitudes we measure
should have a more complex temperature dependence than the relationship indicates.
We have further assumed that the temperature at the time of peak
brightness is the same for all SNe~Ia, which is not true since various SNe~Ia actually exhibit different
temperatures (e.g., Nugent et al. 1995). More specifically, subluminous SNe~Ia like SN 1991bg (e.g.,
Filippenko et al. 1992) tend to have a lower temperature near peak brightness (see also Branch et al. 2006),
while overluminous SNe~Ia like SN 1991T tend to have a higher temperature.
Our assumptions are adopted to motivate the parameterization in Equation~\ref{eq_lvt2pm}; the scatter caused by differing
temperatures will be included in the final dispersion. See Section~\ref{s:goldensample} for
more discussion of the constant-temperature assumption.

Compared to another well-known correlation among SNe~Ia --- the Phillips relation, which shows that the peak
magnitude is correlated with ${\Delta}m_{15}(B)$, the magnitude drop by 15~days after peak time ---
there are three major differences. First, the new relation (Equation~\ref{eq_lvt2pm})
contains three parameters instead of two parameters as in the Phillips relation;
however, similar to ${\Delta}m_{15}(B)$,
the other two parameters in Equation~\ref{eq_lvt2pm} are relatively easy to infer.
Second, Equation~\ref{eq_lvt2pm} focuses on the rising part of the light curves while the Phillips relation involves the declining part.
Third, the new relation has a more straightforward physical explanation (as 
given in the approximate derivation of Equation~\ref{eq_lvt2pm})
compared to the Phillips relation.

In the following section, we will test the new relation by using a sample of SNe~Ia 
having measurements of all three parameters. For comparison purposes, we will also investigate
the relation without considering photospheric velocities, namely
\begin{equation}
M_p = -2.5\,{\rm log}[(t_r)^2] + C,
\label{eq_lt2pm}
\end{equation}
which is equivalent to assuming that the photospheric velocity is the same 
for all SNe~Ia at the time of peak brightness.
We assign the label $M_{t^2}$ to represent $-2.5\,{\rm log}[(t_r)^2] + C$.
In the following analysis, the constant $C$ is assigned a value 0 
for simplicity; we are interested in relative (rather than absolute) trends.

\section{Data Analysis}\label{s:data_analysis}

Zheng et al. (2017) estimated the rise time of 56 SNe~Ia selected from the
well-observed Lick Observatory Supernova Search (LOSS; Filippenko
et al. 2001; Leaman et al. 2011) sample (Ganeshalingam et al. 2010), 
the third Center for Astrophysics sample (CfA3; Hicken et al. 2009a), and 
the Carnegie Supernova Project sample (CSP; Contreras et al. 2010).
While the details are described by Zheng et al. (2017), here we briefly summarize
the rise-time estimation.
The rise time is measured from the well-observed $B$-band light curve with the
first detection being at least 1~mag fainter than the peak magnitude, in order
to measure the first-light time reliably. The $B$ light curve
was fit with a variant broken-power-law function to estimate the first-light time
along with the time of peak brightness to calculate the rise time.
The uncertainty in the rise time is dominated by the estimate of the first-light time, which
includes both the fit-statistic error and the systematic error estimated from
the method.
We start with this sample by excluding the two SNe having redshift
$z < 0.005$ (SN~1999by and SN~2005ke) in order to avoid
large uncertainties in the peculiar velocity, leaving a total of 54 SNe.
Note that we adopted the redshifts $z$ corrected for coherent flows derived from a model
given by Carrick et al. (2015).
We then extract all the necessary information to test the new $M_p \approx M_{v^2t^2}$ relation as
presented in Equation~\ref{eq_lvt2pm},
as well as the $M_p \approx M_{t^2}$ relation of Equation~\ref{eq_lt2pm} for comparison.
For the three parameters in the new $M_p \approx M_{v^2t^2}$ relation
($M_p$, $v_p$, and $t_r$),
we directly adopt the values given by Zheng et al. (2017).

The peak absolute magnitude of each SN is obtained by fitting the peak apparent magnitude,
which was determined when fitting for the time of peak brightness (Zheng et al. 2017),
and correcting for the distance modulus and extinction.
The distance modulus ($\mu$) is calculated from Hubble's law and the measured
value of $z$;
we adopt a standard cosmological model with H$_0 = 70$~km~s$^{-1}$~Mpc$^{-1}$, $\Omega_M = 0.3$, and $\Omega_\Lambda = 0.7$.
The uncertainty in the peak magnitude estimated from 
the observations is usually relatively small ($<0.03$~mag).
We add a residual average peculiar velocity uncertainty of 250~km~s$^{-1}$ 
applied to each SN redshift.
The extinction uncertainties are also considered when calculating the peak absolute magnitude error.
We did not apply $K$-corrections (e.g., Hamuy et al. 1993; Nugent et al. 2002; 
Jha et al. 2007); at such low redshifts (maximum $z = 0.039$),
the typical $K$-correction is very small ($< 0.01$~mag) for SNe at peak
brightness according to Hamuy et al. (1993),
much smaller than the precision discussed in this paper.

An extinction correction was applied to each SN, including
Milky Way Galaxy extinction and host-galaxy extinction. For the 
Galactic extinction, we use the Schlegel et al. (1998) value
rather than the updated Schlafly \& Finkbeiner (2011) value, in order to be consistent with the
MLCS2k2 method,
and adopt $R_V = 3.1$. For the host-galaxy extinction,
we adopt the $A_V$ obtained from MLCS2k2 fitting (Jha et al. 2007), using
$R_V = 1.8$ because there are indications that
$R_V = 3.1$ overestimates host-galaxy extinction (e.g., Hicken et al. 2009b).
However, since the host-galaxy extinction is not well understood, we 
exclude those SNe with host $E(B-V) > 0.3$~mag (or $A_V > 0.54$~mag with 
$R_V = 1.8$).

The photospheric velocity is usually measured from the strong
Si~II $\lambda$6355 absorption line. 
For the purpose of estimating the peak luminosity using our method,
it is best that a good spectrum be taken right at the time of peak brightness,
but this is difficult to do in practice. However, typically the velocity 
is observed to be decreasing linearly around peak brightness 
(e.g., Silverman et al. 2012); hence, as long as there is a spectrum 
observed within a few days of it,
one can extrapolate the velocity to the time of peak brightness. In this paper,
we first adopt the Si~II $\lambda$6355 velocity value from either Silverman et al. 
(2015) or Silverman et al. (2012) if the previous 
one is not available, after -8 days of peak brightness.
Note that in Silverman et al. (2015), two components of the Si~II $\lambda$6355 velocity
are given; one is a high-velocity component and the other is photospheric. Since we are
interested in measuring emission from the photosphere, we adopt the photospheric component.
If a SN has no Si~II $\lambda$6355 value measured as above, we try to collect the data from the
published literature, or we measure it directly from the spectrum found in the public domain.
If more than two data points are measured, we either interpolate or extrapolate.
If only one data point is available, we extrapolate the velocity assuming a velocity gradient of
$-57$~km~s$^{-1}$~day$^{-1}$ as estimated by Silverman et al. (2012); these SNe are labeled in Table 1.
Note that Folatelli et al. (2013) found that the velocity gradient of SNe~Ia can be
quite diverse, ranging from $-42$ to $-250$ km~s$^{-1}$~day$^{-1}$ for different subclasses,
where the normal subclass of SNe~Ia has a velocity gradient of $-86$~km~s$^{-1}$~day$^{-1}$.
To account for the difference from our adopted value of $-57$~km~s$^{-1}$~day$^{-1}$, we add
an additional 50\% uncertainty to the final error estimation for these SNe.

Out of the 54 SNe~Ia in our sample with $z > 0.005$, 7 are excluded because of
their strong host-galaxy extinction of $E(B-V) > 0.3$~mag,
giving us a sample of 47 SNe for the new relation study, which we label ``group1.''
However, all 54 SNe are listed in Table~1 for completeness.
Meanwhile, in order to compare with the Phillips relation, we also 
measure the ${\Delta}m_{15}(B)$ values from the $B$-band light-curve 
fitting (also given by Zheng et al. 2017).

\begin{deluxetable*}{llcccccccccccc}
 \tabcolsep 0.4mm
 \tablewidth{0pt}
 \tablecaption{Full Type Ia Supernova Sample}
  \tablehead{\colhead{SN} & \colhead{Subtype} & \colhead{$z$} & \colhead{$z'^a$} & \colhead{$t_{r,B}$} & \colhead{$v^b$,ref$^c$} & \colhead{Mag$_{p,B}$} & \colhead{AV18$^d$} & \colhead{$\mu$ (mag)} & \colhead{$\mu_{\rm (MLCS)}$} & \colhead{$M_B$\,(mag)} & \colhead{$M_{t^2}$} & \colhead{$M_{v^2t^2}$} & \colhead{group$^e$}}
\startdata
  1998dh  &  Ia-norm  &     0.0077  &     0.0082  &    15.7 &    11.1$\pm0.5$,S12$^f$  &    14.1  &    0.3016  &    32.74  &    32.96  &   -19.34$\pm$0.22  &     -5.97  &    -11.20 & 1 \\
  1998dm  &  Ia-norm  &     0.0055  &     0.0062  &    17.7 &    10.6$\pm0.3$,S15$^f$  &    14.8  &    0.7511  &    32.13  &    33.26  &   -18.68$\pm$0.28  &     -6.24  &    -11.37 & 1 \\
  1999cp  &  Ia-norm  &     0.0103  &     0.0103  &    17.3 &    10.6$\pm0.3$,S15$^f$  &    14.0  &    0.0401  &    33.24  &    33.46  &   -19.36$\pm$0.18  &     -6.19  &    -11.32 & 3 \\
  1999dq  &  Ia-99aa  &     0.0137  &     0.0130  &    18.4 &    10.9$\pm0.1$,S15      &    14.9  &    0.3174  &    33.75  &    33.70  &   -19.82$\pm$0.15  &     -6.32  &    -11.51 & 1 \\
  1999gp  &  Ia-norm  &     0.0260  &     0.0268  &    18.0 &    11.0$\pm0.2$,Z18      &    16.2  &    0.1441  &    35.34  &    35.57  &   -19.61$\pm$0.09  &     -6.27  &    -11.48 & 5 \\
  2000cx  &   Ia-pec  &     0.0070  &     0.0073  &    14.2 &    11.7$\pm0.2$,Z18      &    13.4  &   -0.0520  &    32.47  &    32.64  &   -19.31$\pm$0.24  &     -5.76  &    -11.10 & 2 \\
  2000dn  &  Ia-norm  &     0.0308  &     0.0316  &    16.3 &    10.2$\pm0.2$,S15$^f$  &    16.8  &    0.0191  &    35.71  &    36.13  &   -19.15$\pm$0.07  &     -6.05  &    -11.10 & 5 \\
  2000dr  &  Ia-norm  &     0.0178  &     0.0183  &    12.9 &    10.5$\pm0.3$,S12$^f$  &    16.1  &   -0.0056  &    34.50  &    34.47  &   -18.52$\pm$0.11  &     -5.55  &    -10.65 & 1 \\
  2000fa  &  Ia-norm  &     0.0218  &     0.0223  &    16.7 &    12.0$\pm0.2$,Z18      &    16.1  &    0.2916  &    34.94  &    35.08  &   -19.54$\pm$0.10  &     -6.11  &    -11.51 & 1 \\
  2001en  &  Ia-norm  &     0.0153  &     0.0133  &    16.3 &    12.5$\pm0.4$,S12$^f$  &    15.3  &    0.0729  &    33.79  &    34.36  &   -18.86$\pm$0.15  &     -6.06  &    -11.55 & 1 \\
  2001ep  &  Ia-norm  &     0.0129  &     0.0127  &    16.8 &    10.3$\pm0.3$,S15$^f$  &    15.0  &    0.2551  &    33.69  &    33.89  &   -19.24$\pm$0.15  &     -6.12  &    -11.19 & 1 \\
  2002bo  &  Ia-norm  &     0.0053  &     0.0060  &    15.7 &    13.0$\pm0.3$,S15$^f$  &    14.0  &    0.7674  &    32.05  &    32.32  &   -19.31$\pm$0.29  &     -5.98  &    -11.55 & 1 \\
  2002cr  &  Ia-norm  &     0.0103  &     0.0103  &    16.6 &    10.7$\pm0.3$,S15$^f$  &    14.3  &    0.1776  &    33.24  &    33.44  &   -19.34$\pm$0.18  &     -6.11  &    -11.25 & 3 \\
  2002dj  &  Ia-norm  &     0.0104  &     0.0091  &    15.4 &    14.5$\pm0.3$,Z18      &    14.3  &    0.1341  &    32.98  &    33.22  &   -19.26$\pm$0.21  &     -5.94  &    -11.75 & 1 \\
  2002dl  &   Ia-pec  &     0.0152  &     0.0152  &    13.0 &    10.6$\pm0.4$,Z18$^f$  &    16.1  &   -0.0031  &    34.09  &    34.17  &   -18.28$\pm$0.12  &     -5.58  &    -10.70 & 1 \\
  2002eb  &  Ia-norm  &     0.0265  &     0.0274  &    18.2 &    10.3$\pm0.4$,S15$^f$  &    16.2  &    0.1241  &    35.39  &    35.62  &   -19.60$\pm$0.08  &     -6.30  &    -11.36 & 5 \\
  2002er  &  Ia-norm  &     0.0090  &     0.0092  &    15.9 &    11.8$\pm0.4$,S15$^f$  &    14.8  &    0.3334  &    33.00  &    33.13  &   -19.34$\pm$0.21  &     -6.01  &    -11.37 & 1 \\
  2002fk  &  Ia-norm  &     0.0070  &     0.0073  &    17.9 &     9.8$\pm0.3$,S12$^f$  &    13.3  &    0.0386  &    32.50  &    32.75  &   -19.40$\pm$0.24  &     -6.26  &    -11.22 & 2 \\
  2002ha  &  Ia-norm  &     0.0132  &     0.0135  &    14.9 &    10.8$\pm0.1$,S15      &    15.1  &   -0.0013  &    33.83  &    33.93  &   -19.16$\pm$0.14  &     -5.87  &    -11.04 & 3 \\
  2002he  &  Ia-norm  &     0.0248  &     0.0253  &    13.9 &    12.4$\pm0.1$,S15      &    16.4  &    0.0155  &    35.22  &    35.35  &   -19.01$\pm$0.09  &     -5.71  &    -11.18 & 1 \\
  2003cg  &  Ia-norm  &     0.0053  &     0.0053  &    16.1 &    10.9$\pm0.2$,E06      &    16.0  &    2.2735  &    31.77  &    31.78  &   -19.49$\pm$0.32  &     -6.04  &    -11.22 & 1 \\
  2003fa  &  Ia-99aa  &     0.0391  &     0.0408  &    18.5 &    11.2$\pm0.4$,S15$^f$  &    16.8  &    0.0681  &    36.28  &    36.35  &   -19.80$\pm$0.07  &     -6.34  &    -11.58 & 5 \\
  2003gn  &  Ia-norm  &     0.0333  &     0.0339  &    14.4 &    12.0$\pm0.3$,S15      &    17.5  &    0.1754  &    35.87  &    36.31  &   -18.81$\pm$0.11  &     -5.79  &    -11.19 & 1 \\
  2003gt  &  Ia-norm  &     0.0150  &     0.0154  &    16.7 &    11.1$\pm0.4$,S15$^f$  &    15.4  &    0.1750  &    34.12  &    34.19  &   -19.47$\pm$0.13  &     -6.12  &    -11.34 & 4 \\
   2003W  &  Ia-norm  &     0.0211  &     0.0211  &    14.6 &    14.8$\pm0.4$,S15$^f$  &    16.1  &    0.3353  &    34.81  &    35.03  &   -19.44$\pm$0.10  &     -5.82  &    -11.67 & 1 \\
   2003Y  &  Ia-91bg  &     0.0173  &     0.0170  &    10.5 &     9.8$\pm0.3$,S15$^f$  &    17.9  &    0.5270  &    34.33  &    34.19  &   -17.48$\pm$0.14  &     -5.11  &    -10.06 & 1 \\
  2004at  &  Ia-norm  &     0.0240  &     0.0239  &    16.6 &    11.3$\pm0.1$,Z18      &    15.7  &    0.0185  &    35.09  &    35.31  &   -19.46$\pm$0.09  &     -6.10  &    -11.36 & 5 \\
  2004dt  &  Ia-norm  &     0.0185  &     0.0198  &    16.7 &    13.3$\pm0.3$,S15      &    15.3  &    0.2442  &    34.67  &    34.53  &   -19.83$\pm$0.10  &     -6.11  &    -11.73 & 1 \\
  2004ef  &  Ia-norm  &     0.0298  &     0.0301  &    13.8 &    12.8$\pm0.4$,S15$^f$  &    17.1  &    0.1771  &    35.60  &    35.75  &   -19.03$\pm$0.10  &     -5.70  &    -11.23 & 1 \\
  2004eo  &  Ia-norm  &     0.0148  &     0.0154  &    15.8 &    10.8$\pm0.4$,S15$^f$  &    15.5  &    0.1383  &    34.12  &    34.10  &   -19.29$\pm$0.14  &     -5.99  &    -11.16 & 4 \\
  2005cf  &  Ia-norm  &     0.0070  &     0.0071  &    15.9 &    10.1$\pm0.1$,S15      &    13.7  &    0.1612  &    32.42  &    32.65  &   -19.41$\pm$0.25  &     -6.01  &    -11.03 & 2 \\
  2005de  &  Ia-norm  &     0.0149  &     0.0151  &    17.4 &    10.2$\pm0.2$,S15$^f$  &    15.8  &    0.2376  &    34.08  &    34.51  &   -19.06$\pm$0.13  &     -6.20  &    -11.25 & 1 \\
  2005ki  &  Ia-norm  &     0.0203  &     0.0200  &    15.1 &    11.1$\pm0.2$,S15$^f$  &    15.6  &    0.0169  &    34.69  &    34.85  &   -19.20$\pm$0.10  &     -5.89  &    -11.12 & 4 \\
   2005M  &   Ia-91T  &     0.0230  &     0.0255  &    20.5 &    10.6$\pm0.2$,S15$^f$  &    16.0  &    0.1877  &    35.24  &    35.50  &   -19.68$\pm$0.09  &     -6.56  &    -11.68 & 1 \\
  2006cp  &  Ia-norm  &     0.0233  &     0.0222  &    17.5 &    13.6$\pm0.3$,Z18      &    16.0  &    0.1746  &    34.92  &    35.34  &   -19.27$\pm$0.11  &     -6.22  &    -11.88 & 1 \\
  2006gr  &  Ia-norm  &     0.0335  &     0.0342  &    18.3 &    11.4$\pm0.3$,Z18      &    17.3  &    0.2763  &    35.89  &    36.29  &   -19.41$\pm$0.09  &     -6.31  &    -11.59 & 1 \\
  2006le  &  Ia-norm  &     0.0172  &     0.0189  &    16.6 &    11.6$\pm0.3$,Z18      &    16.4  &    0.0234  &    34.57  &    34.78  &   -19.85$\pm$0.11  &     -6.10  &    -11.42 & 4 \\
   2006X  &  Ia-norm  &     0.0064  &     0.0059  &    16.0 &    14.7$\pm0.1$,S15      &    15.5  &    2.3994  &    32.03  &    31.11  &   -20.34$\pm$0.30  &     -6.02  &    -11.85 & 1 \\
  2007af  &  Ia-norm  &     0.0062  &     0.0062  &    16.8 &    10.6$\pm0.1$,S15      &    13.3  &    0.2198  &    32.14  &    32.34  &   -19.31$\pm$0.28  &     -6.12  &    -11.25 & 1 \\
  2007le  &  Ia-norm  &     0.0067  &     0.0063  &    15.0 &    14.0$\pm0.4$,S15$^f$  &    14.0  &    0.5883  &    32.18  &    32.57  &   -19.24$\pm$0.27  &     -5.88  &    -11.61 & 1 \\
  2007qe  &  Ia-norm  &     0.0244  &     0.0195  &    16.1 &    14.1$\pm0.3$,Z18      &    16.2  &    0.1741  &    34.64  &    35.47  &   -18.91$\pm$0.11  &     -6.03  &    -11.78 & 1 \\
  2008bf  &  Ia-norm  &     0.0251  &     0.0271  &    16.7 &    11.5$\pm0.2$,F13      &    15.8  &   -0.0104  &    35.37  &    35.47  &   -19.66$\pm$0.07  &     -6.11  &    -11.42 & 5 \\
  2008ec  &  Ia-norm  &     0.0149  &     0.0158  &    15.5 &    10.5$\pm0.1$,S15      &    15.8  &    0.3599  &    34.18  &    34.31  &   -19.21$\pm$0.13  &     -5.96  &    -11.06 & 1 \\
\\                                                                                     
   2001V  &  Ia-norm  &     0.0162  &     0.0156  &    17.0 &    11.6$\pm0.3$,Z18$^f$  &    14.7  &    0.1436  &    34.15  &    34.10  &   -19.70$\pm$0.13  &     -6.15  &    -11.47 & 4 \\
  2005hk  &      Iax  &     0.0118  &     0.0118  &    18.1 &     5.7$\pm0.3$,P07      &    15.9  &    0.6691  &    33.53  &    34.62  &   -18.74$\pm$0.16  &     -6.28  &    -10.06 & 1 \\
  2006ax  &  Ia-norm  &     0.0180  &     0.0179  &    18.2 &    10.5$\pm0.1$,Z18      &    15.2  &    0.0097  &    34.45  &    34.73  &   -19.44$\pm$0.11  &     -6.30  &    -11.41 & 4 \\
  2006lf  &  Ia-norm  &     0.0130  &     0.0120  &    14.8 &    11.4$\pm0.4$,S15$^f$  &    17.7  &    0.0290  &    33.58  &    33.99  &   -19.39$\pm$0.16  &     -5.86  &    -11.14 & 3 \\
  2007bd  &  Ia-norm  &     0.0319  &     0.0317  &    13.5 &    12.8$\pm0.4$,S15$^f$  &    16.7  &   -0.0012  &    35.72  &    35.87  &   -19.16$\pm$0.07  &     -5.65  &    -11.18 & 1 \\
  2007ci  &  Ia-norm  &     0.0194  &     0.0178  &    13.0 &    11.8$\pm0.2$,S15      &    15.9  &   -0.0072  &    34.44  &    34.55  &   -18.58$\pm$0.11  &     -5.57  &    -10.93 & 1 \\
\\                                                                                     
  2005kc  &  Ia-norm  &     0.0137  &     0.0145  &    17.4 &    10.6$\pm0.3$,F13      &    16.0  &    0.6070  &    33.99  &    34.21  &   -19.42$\pm$0.16  &     -6.20  &    -11.32 & 1 \\
  2007on  &  Ia-norm  &     0.0062  &     0.0074  &    14.8 &    11.3$\pm0.3$,S15$^f$  &    13.1  &   -0.0447  &    32.52  &    31.54  &   -19.41$\pm$0.23  &     -5.85  &    -11.11 & 1 \\
  2008bc  &  Ia-norm  &     0.0157  &     0.0156  &    16.8 &    11.6$\pm0.2$,F13      &    15.7  &   -0.0142  &    34.15  &    34.89  &   -19.52$\pm$0.12  &     -6.13  &    -11.45 & 4 \\
  2008gp  &  Ia-norm  &     0.0328  &     0.0338  &    17.5 &    11.3$\pm0.4$,G08$^f$  &    16.9  &    0.0346  &    35.86  &    35.93  &   -19.50$\pm$0.08  &     -6.22  &    -11.48 & 5 \\
  2008hv  &  Ia-norm  &     0.0125  &     0.0140  &    14.4 &    10.9$\pm0.2$,F13      &    14.9  &    0.0002  &    33.91  &    34.07  &   -19.13$\pm$0.13  &     -5.80  &    -10.99 & 3 \\
\enddata
\tablenotetext{a}{Flow corrected.}
\tablenotetext{b}{Measured from ${\rm Si~II}~\lambda6355$, and in units of k~km~s$^{-1}$.}
\tablenotetext{c}{Detailed references: E06, Elias-Rosa et al. 2006; P07, Phillips et al. 2007; G08, Garnavich 2008; S12, Silverman et al. 2012; F13, Folatelli et al. 2013; S15, Silverman et al. 2015; Z18, this work.}
\tablenotetext{d}{$A_V$ from MLCS2k2 fitting with $R_V = 1.8$.}
\tablenotetext{e}{Larger-number groups are included in the smaller-number groups.}
\tablenotetext{f}{Used the extrapolation method; see text for details.}
\end{deluxetable*}

\section{Results: The New Relation}\label{s:relation}
\subsection{Full Sample}\label{s:relation_full_sample}
\begin{figure*}
\includegraphics[width=.320\textwidth]{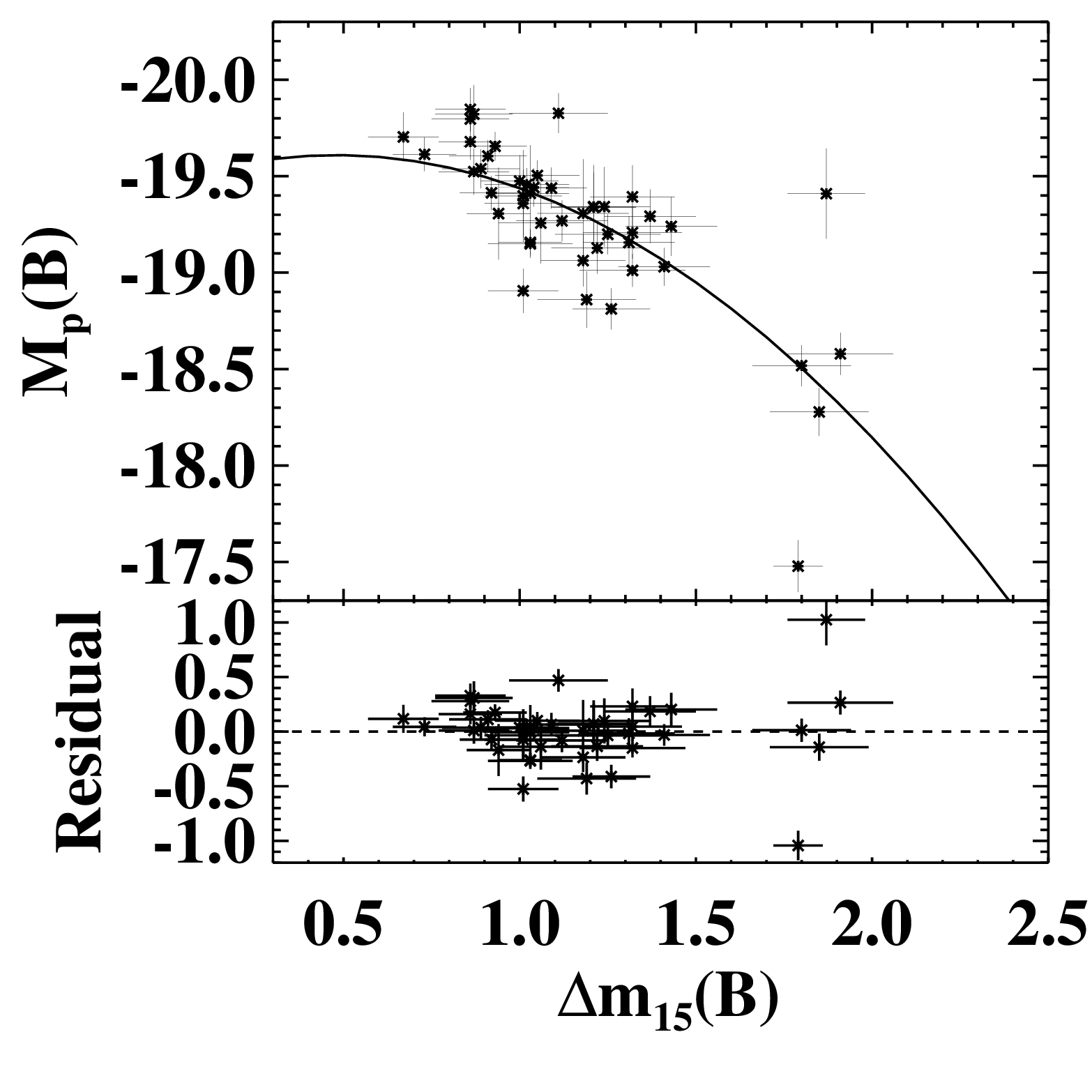}
\includegraphics[width=.320\textwidth]{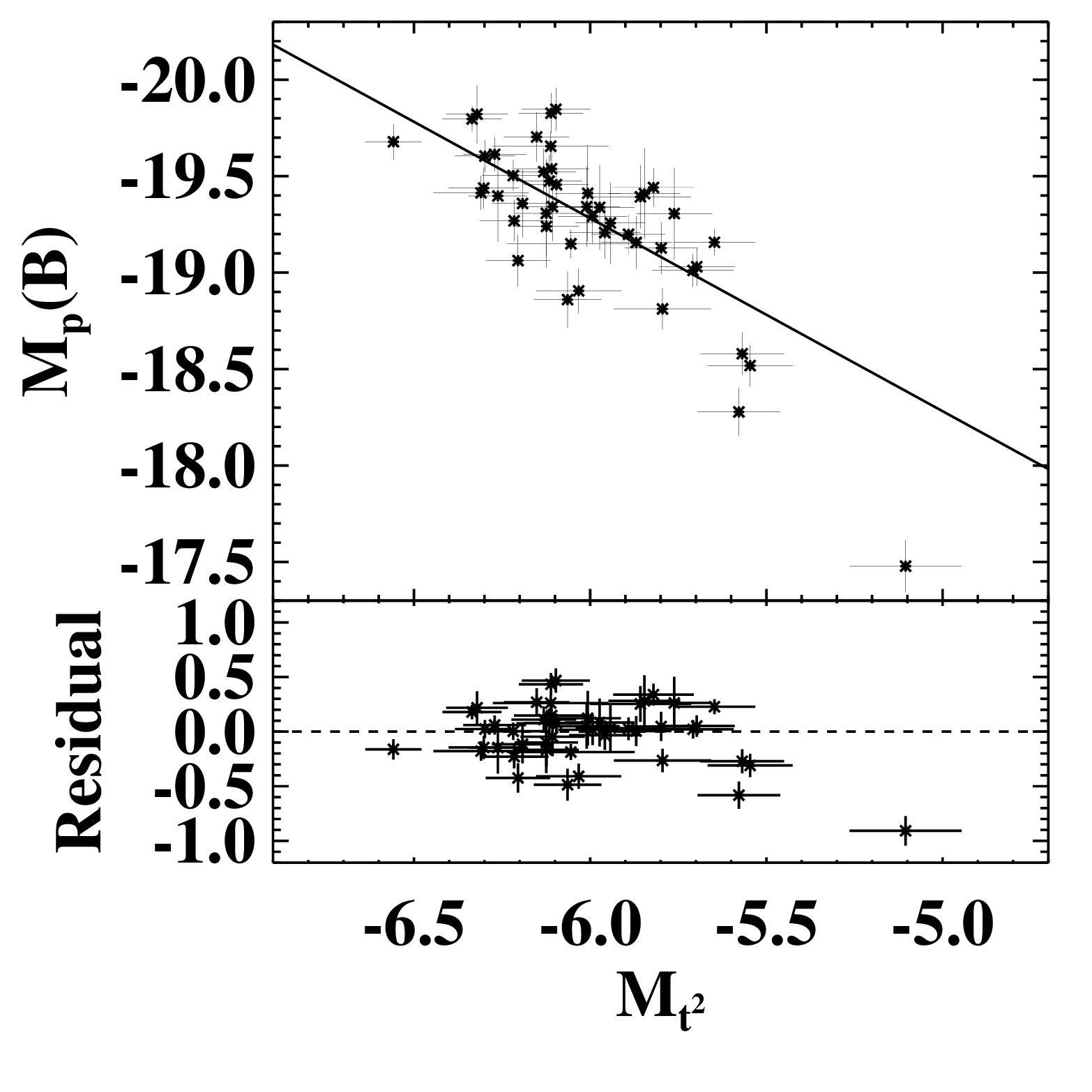}
\includegraphics[width=.320\textwidth]{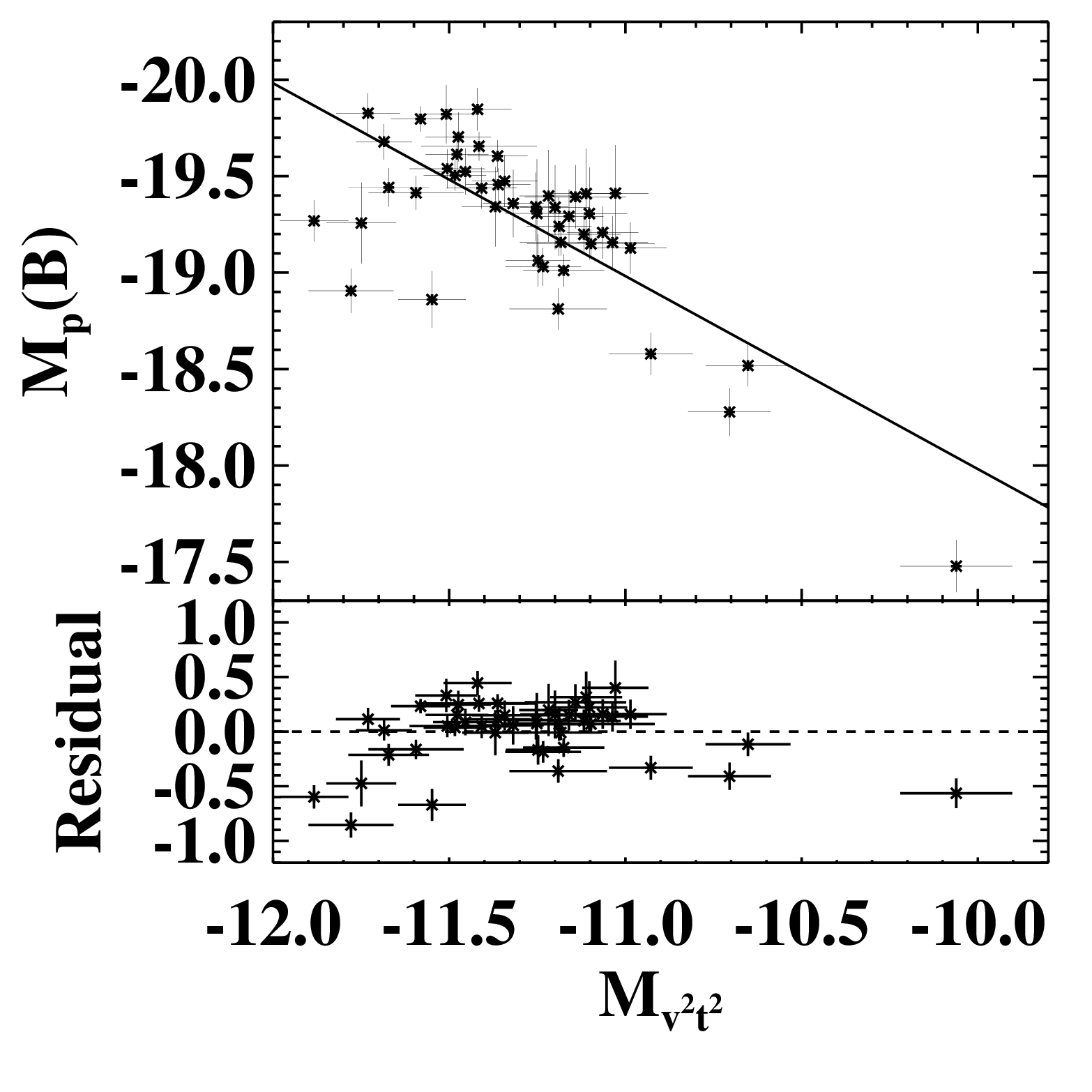}
\caption{The new three-parameter relation ($M_p$ vs. $M_{v^2t^2}$ from Equation~\ref{eq_lvt2pm}, right panel) compared with the
         the Phillips relation ($M_p$ vs. ${\Delta}m_{15}(B)$, left panel) and 
  the relation without considering photospheric velocities
	 ($M_p$ vs. $M_{t^2}$ from Equation~\ref{eq_lt2pm}, middle panel) 
   for the ``group1'' sample.}
\label{relation_big_sample}
\end{figure*}

First, we plot the Phillips relation ($M_p$ vs. ${\Delta}m_{15}(B)$)
derived from our sample in the left panel of Figure~\ref{relation_big_sample}.
The three-parameter new relation (Equation~\ref{eq_lvt2pm}; $M_p$ vs. $M_{v^2t^2}$)
is shown in the right panel of Figure~\ref{relation_big_sample}, and
for comparison, the relation without considering photospheric velocities 
(Equation~\ref{eq_lt2pm}; $M_p$ vs. $M_{t^2}$) is in the middle panel.
All three panels exhibit clear relations
between the peak magnitude and the corresponding abscissa values.

To find the best fit for all three relations, we use an IDL implementation of
{\it mpfit} (Markwardt 2009)\footnote{https://www.physics.wisc.edu/$^\sim$craigm/idl/fitting.html}
to fit each dataset in the three panels with slightly different procedures. For the
Phillips relation (left panel), we use the quadratic function given by Phillips et al. (1999);
however, we fix both parameters $a$ and $b$ with the values given in Table~3 of Phillips et al. (1999)
for the $B$ band except for the ordinate-axis constant (i.e., we fit only the ordinate-axis shift to match our data).
We also performed fitting where all parameters were free, and the results are shown in Table 3.
For the Phillips relation, except for the ``group1'' case, the AIC$_c$ (see below for the definition) 
finds no strong evidence in favor of allowing all parameters to be free ($k=3$).

For the other two relations, a linear function is adopted to
find the best fit (as shown by Equations~\ref{eq_lt2pm} and \ref{eq_lvt2pm}),
but with the slope fixed to be 1.0. Specifically, we also only fit the ordinate-axis shift.
A slope of unity is expected naively for the expanding fireball model near peak (Equations~\ref{eq_lvt2pm} and \ref{eq_lt2pm}), and, when fitting instead with all parameters free, we find the best-fit result to be $M_p=1.05(\pm0.06) \times M_{v^2t^2} -7.37$.
The fitting results are shown in Figure~\ref{relation_big_sample}.
Model comparison (see Table~3) provides no evidence in favor of adding the additional parameters to the models. 

In Table~2, we list the $\chi^2$ statistics and Hubble residual scatter for these different 
relations, as well as those calculated from MLCS2k2 distances.
The peak absolute magnitude of the SN is the only dependent variable used to calculate $\chi^2$.
We perform model comparisons by measuring the $\chi^2$ values for the relations and use the Akaike Information Criterion (AIC; Akaike 1974) to apply a penalty according to the number of fit parameters and number of data points.
MLCS2k2 estimates distance moduli $\mu_{\rm MLCS2k2}$ to individual SNe, and we compute residuals from the relation $\mu_{\rm MLCS2k2} = 5 \log(z) + b_{\rm MLCS2k2}$, where we allow
$b_{\rm MLCS2k2}$ to vary during a fit to all SN distances in each sample.

\begin{deluxetable*}{c c|c|c|c|c|c|c}
 \tabcolsep 0.4mm
 \tablewidth{0pt}
 \tablecaption{Fitting Results}
  \tablehead{\colhead{Groups} & \colhead{Cut}       & \colhead{Values} & \colhead{Hubble residual             } & \colhead{Phillips         } & \colhead{$M_p$ vs. $M_{t^2}$         } & \colhead{$M_p$ vs. $M_{v^2t^2}$         } & \colhead{Peculiar scatter     } \\
             \colhead{      } & \colhead{         } & \colhead{      } & \colhead{                   (MLCS2k2)} & \colhead{         relation} & \colhead{                   relation} & \colhead{                      relation} & \colhead{            250 km s$^{-1}$} }
\startdata
\hline
1      & $z > 0.005$            & $\chi^2$/dof        &158.83/46=3.45   & 240.00/46=5.22  & 239.25/46=5.20  &  263.27/46=5.72   &  \\
full   & $E(B-V) < 0.3$         & residual scatter    & 0.263$\pm$0.058 & 0.292$\pm$0.054 & 0.268$\pm$0.035 &  0.290$\pm$0.035  & 0.136$\pm$0.021 \\
$N=47$   & no $v_{\rm Si~II}$ cut & ${\Delta}$AIC$_c^a$ & -104.44         & -23.27          & -24.02          &  0                &  \\
\hline
2      & $z > 0.005$            & $\chi^2$/dof        & 46.66/21=2.22   & 47.42/21=2.26   & 47.17/21=2.25   &  19.35/21=0.92    &  \\
low-$v$ golden      & $E(B-V) < 0.1$         & residual scatter    & 0.165$\pm$0.040 & 0.143$\pm$0.021 & 0.161$\pm$0.024 &  0.108$\pm$0.018  & 0.135$\pm$0.030 \\
$N=22$   & $v_{\rm Si~II} < 12.0$ & ${\Delta}$AIC$_c$     & 27.31           & 28.07           & 27.82           &  0                &  \\
\hline
\hline
3      & $z > 0.010$            & $\chi^2$/dof        & 46.58/18=2.59   & 46.11/18=2.56   & 45.65/18=2.54   &  18.40/18=1.02    &  \\
low-$v$ golden     & $E(B-V) < 0.1$         & residual scatter    & 0.181$\pm$0.038 & 0.143$\pm$0.023 & 0.160$\pm$0.029 &  0.102$\pm$0.021  & 0.107$\pm$0.022 \\
$N=19$   & $v_{\rm Si~II} < 12.0$ & ${\Delta}$AIC$_c$     & 28.18           & 27.71           & 27.25           &  0                &  \\
\hline
4      & $z > 0.015$            & $\chi^2$/dof        & 44.41/13=3.42   & 41.49/13=3.19   & 42.47/13=3.27   &  17.46/13=1.34    &  \\
low-$v$ golden      & $E(B-V) < 0.1$         & residual scatter    & 0.205$\pm$0.046 & 0.143$\pm$0.033 & 0.168$\pm$0.034 &  0.111$\pm$0.024  & 0.086$\pm$0.020 \\
$N=14$   & $v_{\rm Si~II} < 12.0$ & ${\Delta}$AIC$_c$     & 26.95           & 24.03           & 25.01           &  0                &  \\
\hline
5      & $z > 0.020$            & $\chi^2$/dof        & 16.46/6 =2.74   & 34.08/6 =5.68   & 22.72/6 =3.79   &   8.61/6 =1.43    &  \\
low-$v$ golden      & $E(B-V) < 0.1$         & residual scatter    & 0.121$\pm$0.036 & 0.170$\pm$0.056 & 0.142$\pm$0.037 &  0.091$\pm$0.015  & 0.060$\pm$0.018 \\
$N=7$    & $v_{\rm Si~II} < 12.0$ & ${\Delta}$AIC4$_c$     & 7.85            & 25.47           & 14.11           &  0                &  \\
\hline
\enddata
\tablenotetext{a}{${\Delta}$AIC$_c$ is compared to the $M_p$ vs. $M_{v^2t^2}$ relation. Larger difference of ${\Delta}$AIC$_c$ ($> 6$) offers strong positive evidence, namely the $M_p$ vs. $M_{v^2t^2}$ relation is much better. From this table one can see that except group1 compared to MLCS2k2, the $M_p$ vs. $M_{v^2t^2}$ relation shows strong improvement to nearly all the cases.}
\end{deluxetable*}

The AIC makes it possible to 
perform model selection when the models have different numbers of free parameters. 
Here we use AIC$_c$ = $\chi^2 + 2k + \frac{2k(k+1)}{N - k -1}$
(where $k$ is the number of parameters and $N$ is the number of data points used in fit),
which is a version of the AIC corrected
for small datasets (Sugiura 1978). 
The AIC$_c$ penalizes $\chi^2$ for the number of free parameters.
A difference of 2 in the AIC$_c$ provides positive evidence for the model having lower AIC$_c$, 
while a difference of 6 offers strong positive evidence (e.g., Kass \& Raftery 1995; Mukherjee et al. 1998). 
For all of the relations, the peak absolute magnitude of the SN is the only dependent variable used to calculate $\chi^2$, and
the number of free parameters $k$ is equal to one. 

\begin{figure*}
\includegraphics[width=.320\textwidth]{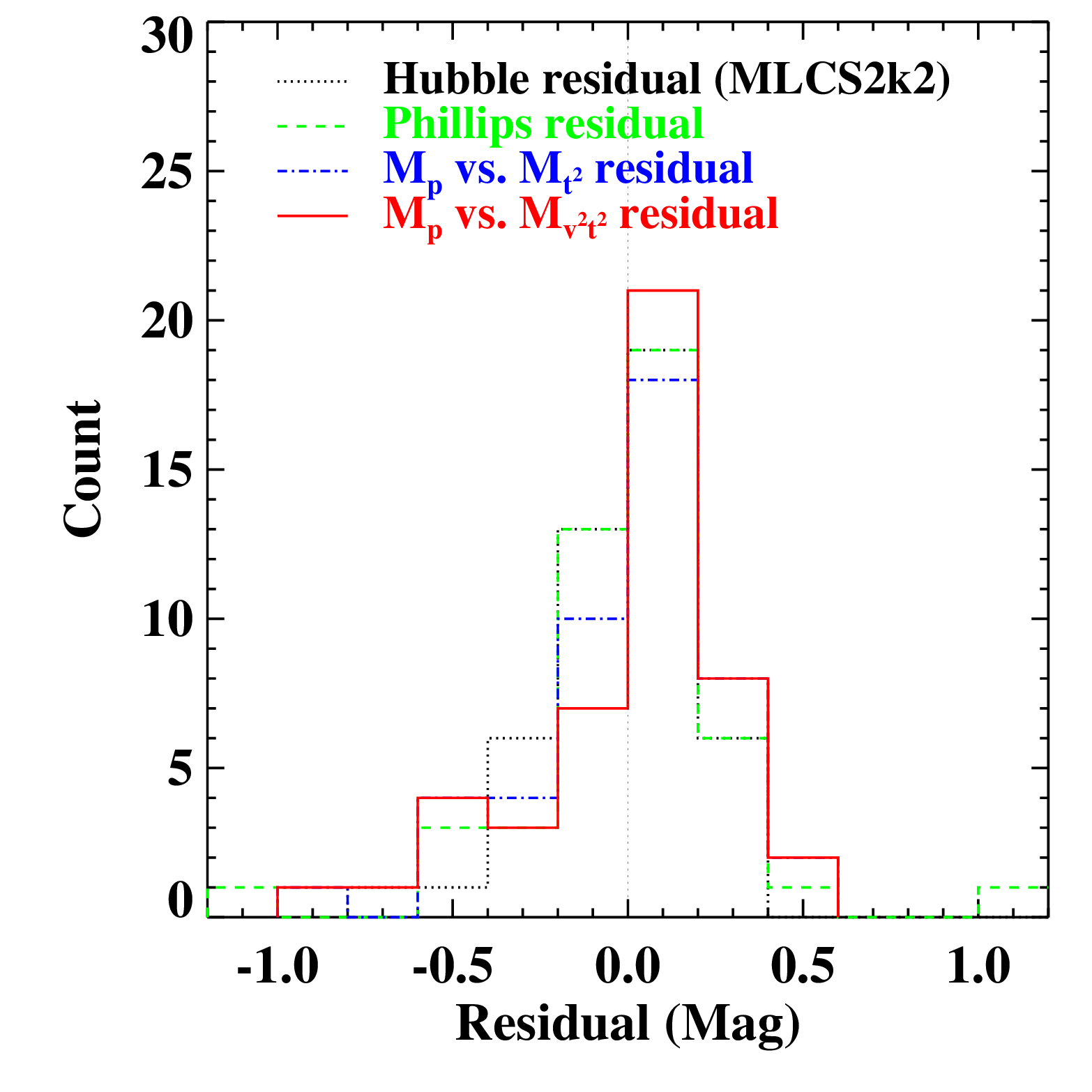}
\includegraphics[width=.320\textwidth]{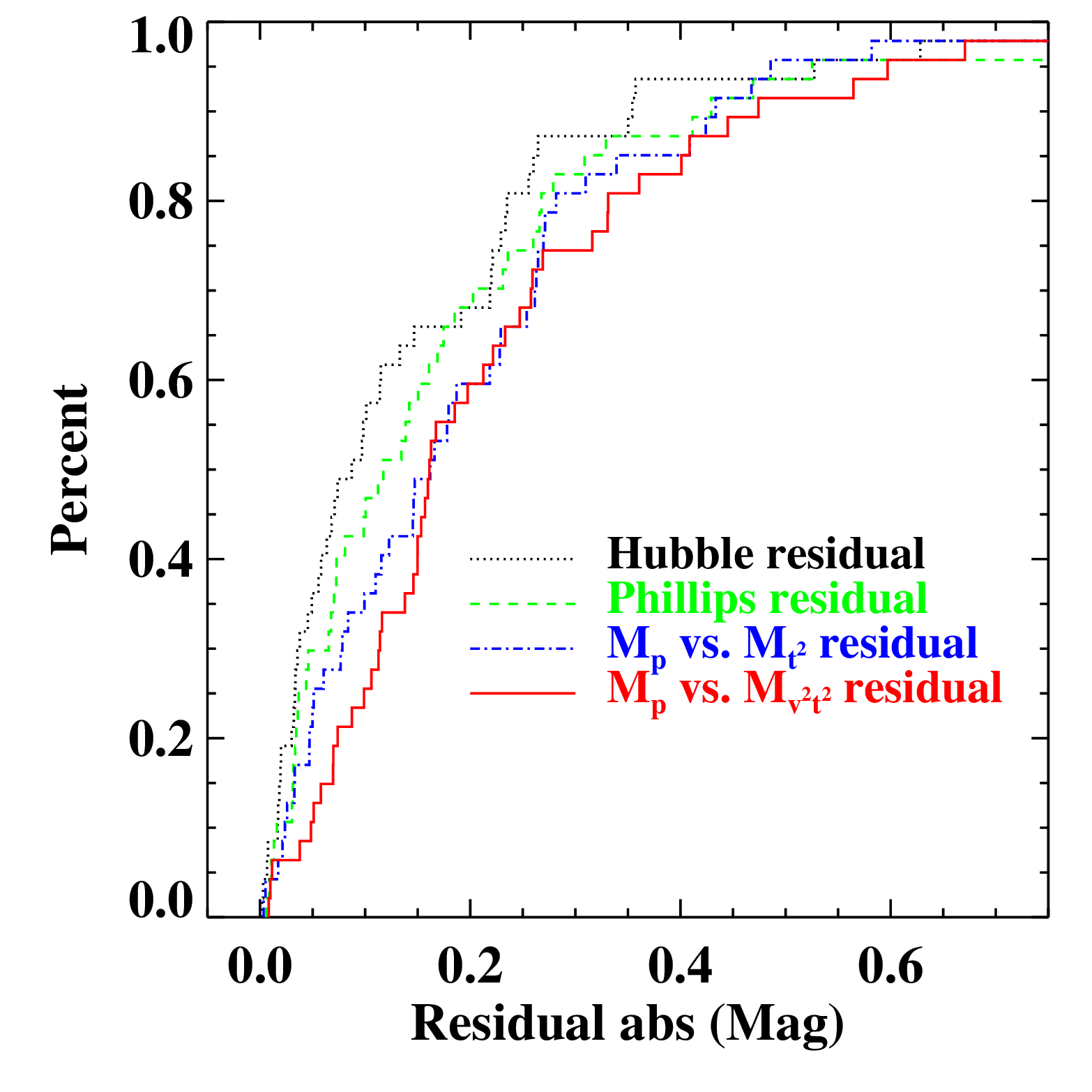}
\includegraphics[width=.320\textwidth]{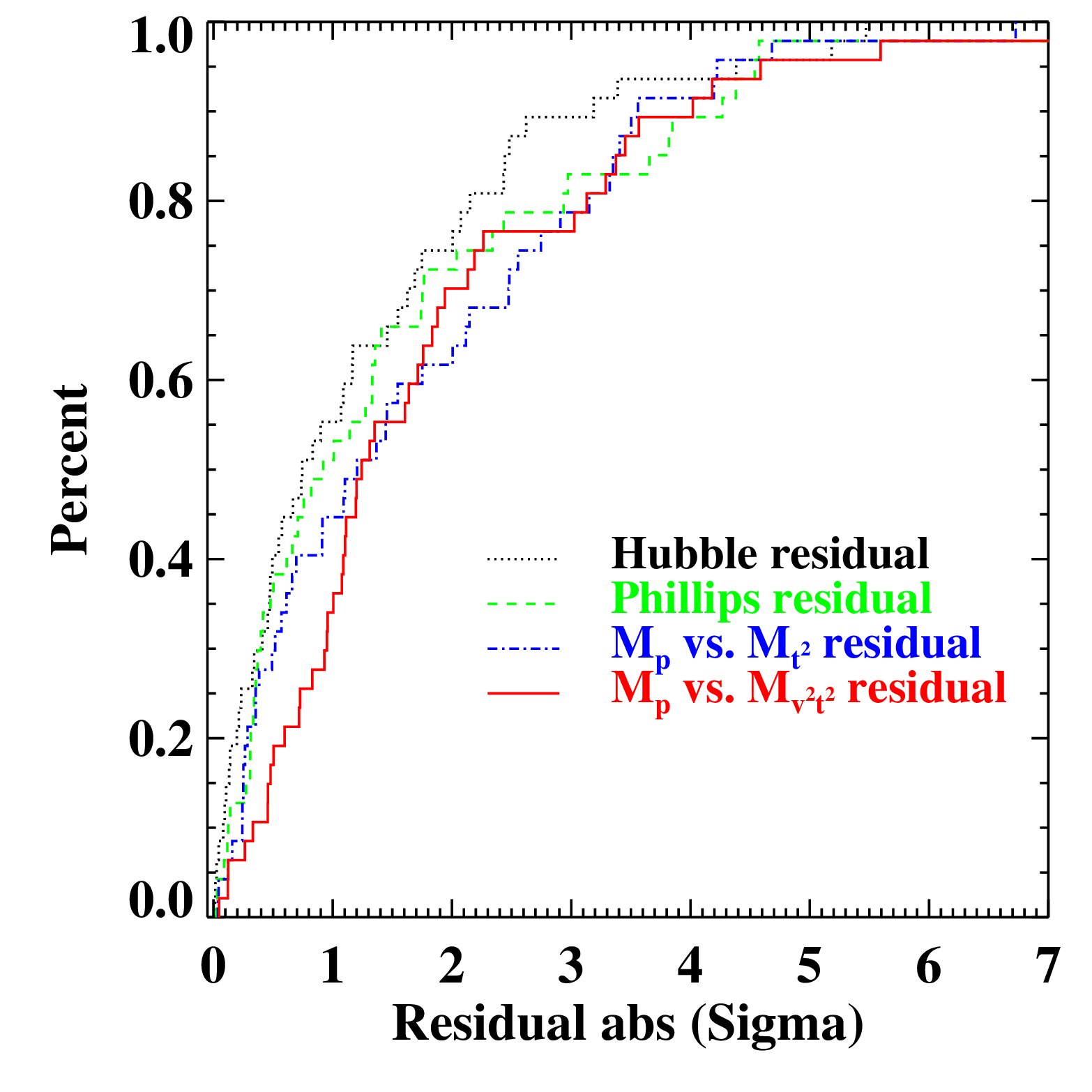}
\caption{Histogram (left panel) and cumulative (middle panel) distributions of
         the residuals, and the absolute value of the residual (residual divided by the
         uncertainty; right panel) distribution for the four cases (the three relations shown
	 in Figure~\ref{relation_big_sample} plus Hubble residuals from 
         MLCS2k2 fitting) for ``group1.''
	 The residuals are generally comparable.}
\label{relation_residual_histgram}
\end{figure*}

In the left panel of Figure~\ref{relation_residual_histgram},
we show the histogram distributions of the residuals from ``group1'' for all 
four cases (the three relations shown in the Figure~\ref{relation_big_sample} plus
the Hubble residual from MLCS2k2 fitting), 
while the middle and right panel shows the corresponding cumulative distributions.
All three relations have comparable scatter in their residuals, with the MLCS2k2 residuals having slightly smaller dispersion.
We calculate the $1\sigma$ scatter of the Hubble residual from
MLCS2k2 fitting to be $0.263 \pm 0.058$~mag,
for the Phillips relation to be $0.292 \pm 0.054$~mag, for the 
$M_p$ vs. $M_{t^2}$ relation to be $0.268 \pm 0.035$~mag,
and for the $M_p$ vs. $M_{v^2t^2}$ relation to be $0.290 \pm 0.035$~mag. 
Here the uncertainties in the residual scatter for the three relations are 
estimated using a bootstrap procedure:
from a sample of $N$ SNe, we randomly pick a SN and repeat $N$ times to obtain
an $N$ SN sample (some of the SNe may be picked more than once), we fit this 
sample with the same method as performed for the three relations shown 
in Figure~\ref{relation_residual_histgram} and calculate the residual for 
each one, the procedure is repeated 1000 times,
and the scatter is then calculated.

Since all models have the same number of free parameters and will be equally penalized by the AIC$_c$, we can directly compare $\chi^2$ values.
Table 2 shows that for ``group1,'' the $M_p$ vs. $M_{v^2t^2}$ relation yields a worse model than the $M_p$ vs. $M_{t^2}$ relation and the Phillips relation. The Hubble residual from MLCS2k2 fitting yields the smallest $\chi^2$ value by a significant margin, which means the MLCS2k2 fitting is the best method for this ``group1'' sample.
A caveat about using $\chi^2$ and AIC$_c$ for comparison is that both of these statistical measures assume that the data are drawn from a normal distribution, and are therefore sensitive to outliers and skewed distributions, which are seen in the ``group1'' sample. We also list the residual scatter in Table 2 as an additional parameter for comparison.

\begin{figure*}
\includegraphics[width=.320\textwidth]{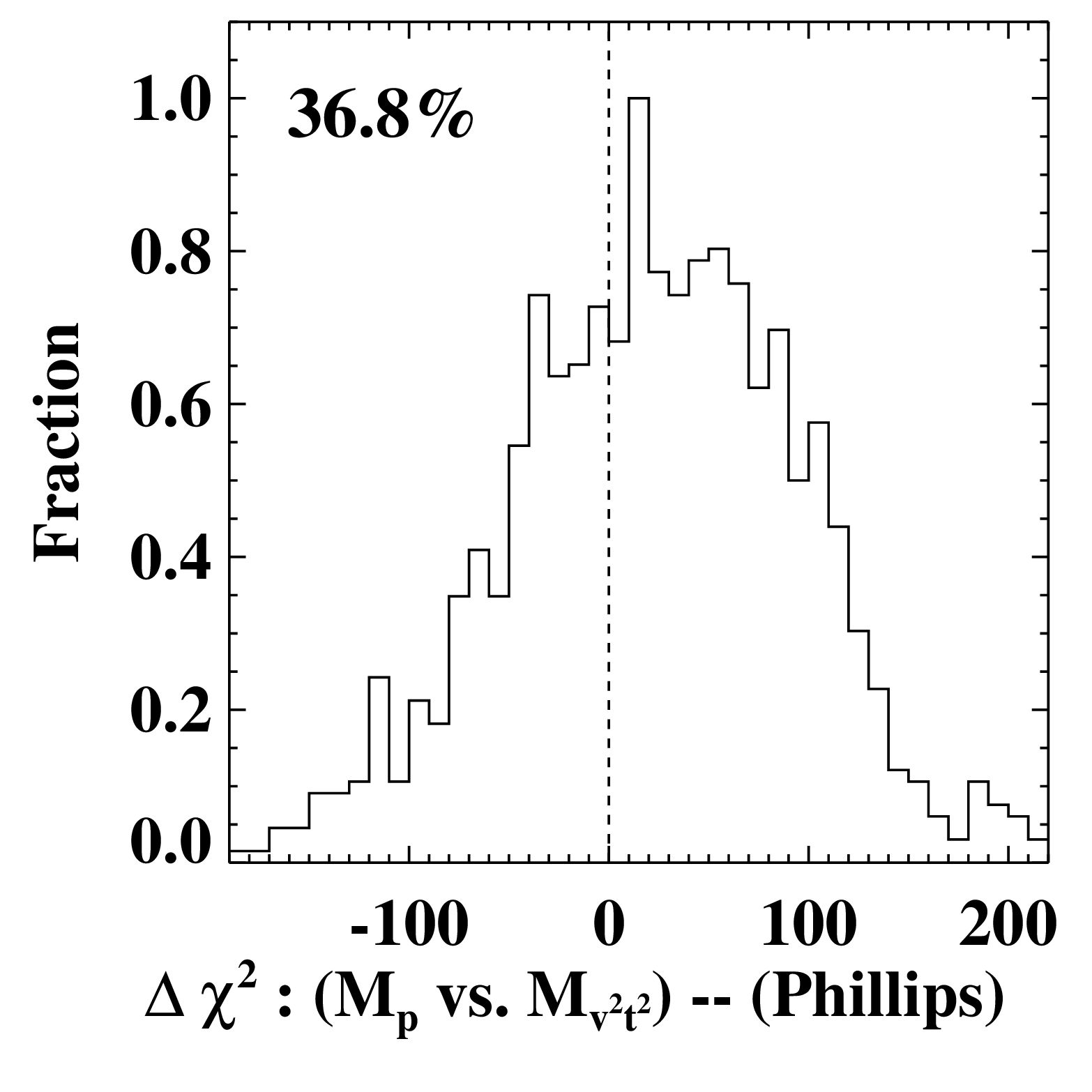}
\includegraphics[width=.320\textwidth]{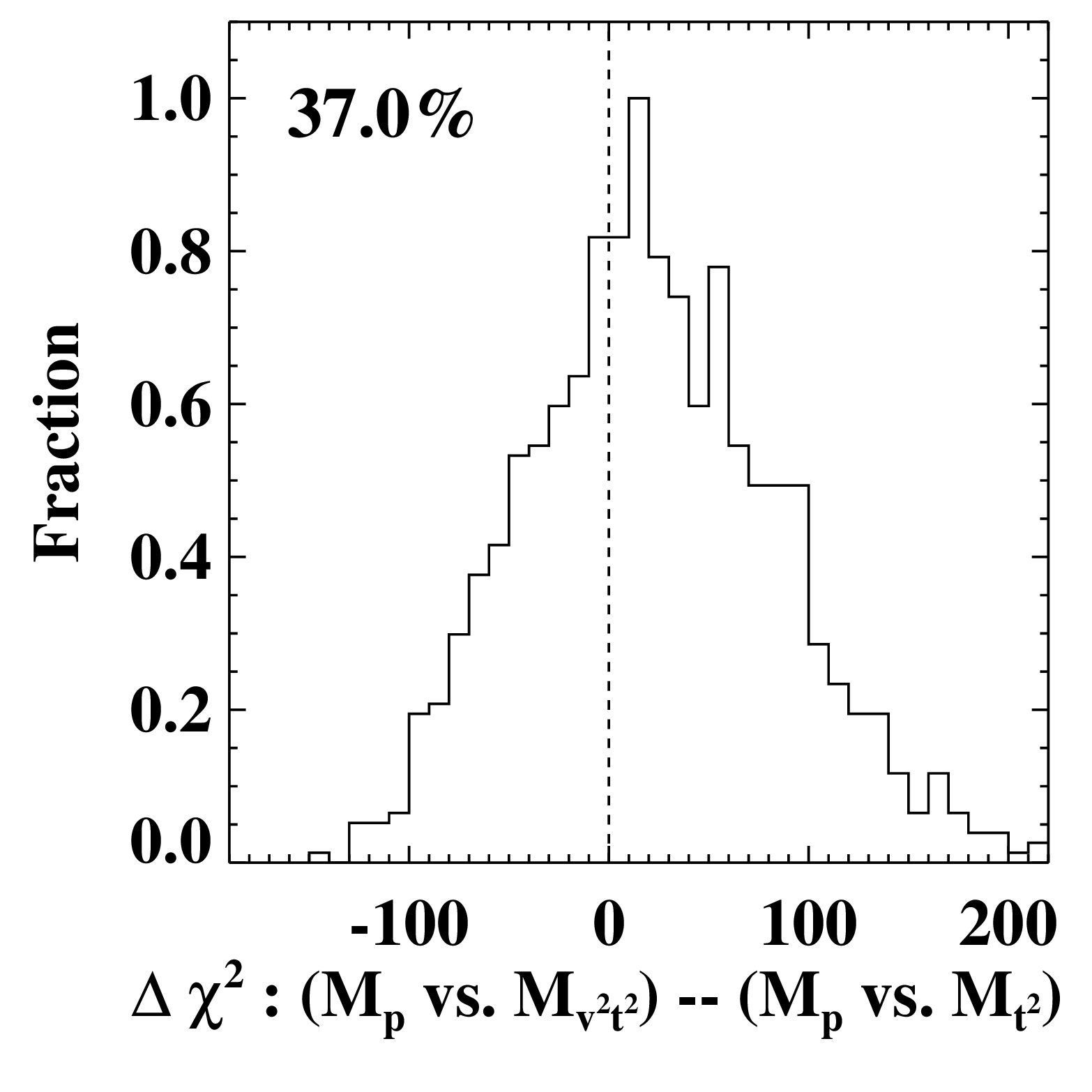}
\includegraphics[width=.320\textwidth]{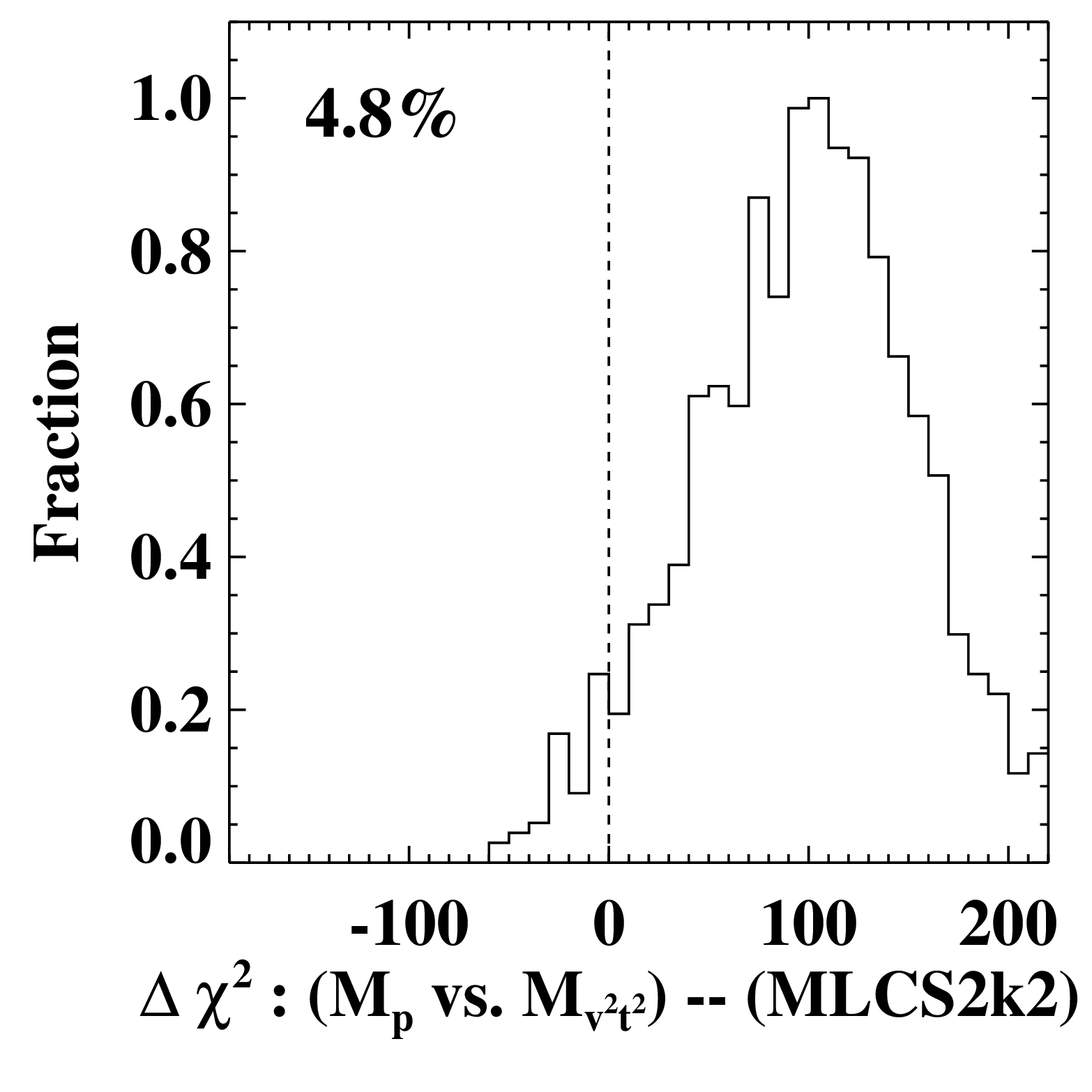}
\caption{Histogram distribution of the $\chi^2$ difference between the 
  $M_p$ vs. $M_{v^2t^2}$ relation and the
         others, from bootstrap simulations (see text for details). 
   A residual difference less than zero
	 indicates that the $M_p$ vs. $M_{v^2t^2}$ relation is better than 
   the comparison.
	 For 1000 simulations, the statistical probability (given
   by the percentage in
	 each panel) shows that the $M_p$ vs. $M_{v^2t^2}$
	 relation is clearly better than the Phillips relation 
    and the $M_p$ vs. $M_{t^2}$ relation,
	 and it is essentially as good as the MLCS2k2 fitting method.}
\label{relation_result_bootstrap_group1}
\end{figure*}

To additionally compare the four light-curve models, we examine $\chi^2$ from the above bootstrap procedure fitting.
For 1000 simulations, we compare $\chi^2$ from
the $M_p$ vs. $M_{v^2t^2}$ relation fitting with the other three cases 
by subtracting $\chi^2$ from each other\footnote{It is only appropriate 
to do this when the fitting parameter number is the same for each 
relation fitting, which applies to our cases; see below for more 
discussion.}
and plot the histogram distribution for the differences. This is shown in
Figure~\ref{relation_result_bootstrap_group1}, where a $\chi^2$ difference less than zero indicates that
the $M_p$ vs. $M_{v^2t^2}$ relation's $\chi^2$ is smaller than that of the comparison model. 
For 1000 simulations, $M_p$ vs. $M_{v^2t^2}$ only has a smaller $\chi^2$ than the Phillips 
relation, the $M_p$ vs. $M_{t^2}$ relation, and the MLCS2k2 fitting
for 36.8\%, 37.0\%, and 4.8\% (respectively) of the bootstrap realizations. This shows that, for the ``group1'' sample, MLCS2k2 fitting is likely the best method.

Although the $M_p$ vs. $M_{v^2t^2}$ relation does not show improvements compared to the other three methods, they may have ($M_p$ vs. $M_{t^2}$ and $M_p$ vs. $M_{v^2t^2}$) a linear functional form.
Unlike the Phillips relation, which becomes nonlinear at the faint end where most objects are subluminous SNe~Ia like SN 1991bg (e.g., Filippenko et al. 1992) having large ${\Delta}m_{15}(B)$ values, the new relation is linear throughout the entire abscissa range. This can be directly read from  Equation~\ref{eq_lvt2pm}, which also gives a probable physical explanation for the new relation.

\subsection{Subgroups from the Full Sample}

In the above analysis, we selected SNe~Ia with very loose criteria; the 
only two cuts were to exclude
objects at $z < 0.005$ and objects with host-galaxy extinction $E(B-V) > 0.3$~mag. 
Application of tighter constraints could
likely reduce the scatter. For example, SNe~Ia with large 
${\Delta}m_{15}(B)$ are thought to be outliers from the Phillips relation,
and should therefore be excluded. Also, a more stringent extinction
cut could improve the fit. Here, we study several subgroups from the 
full sample and examine their properties.

\begin{figure*}
\includegraphics[width=.320\textwidth]{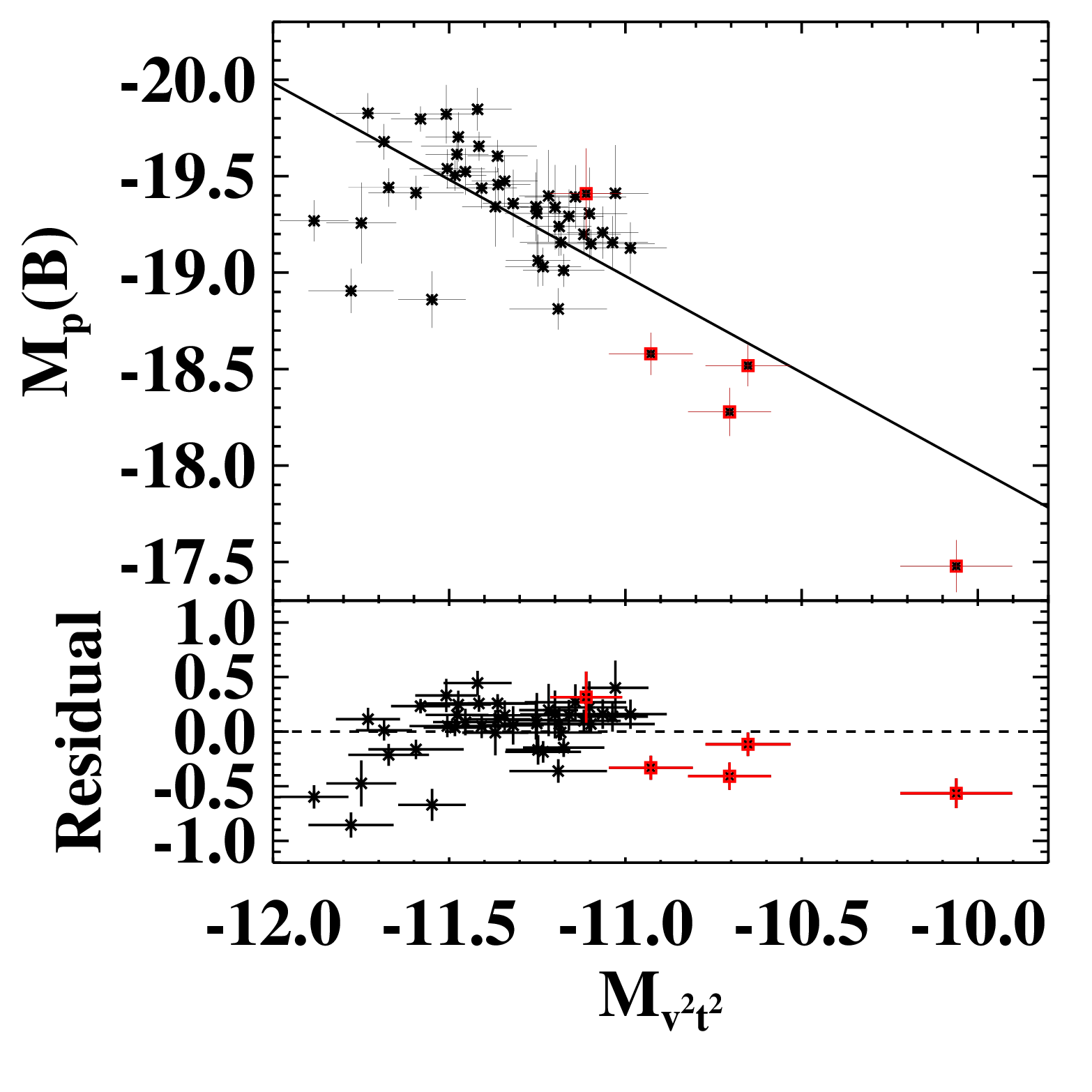}
\includegraphics[width=.320\textwidth]{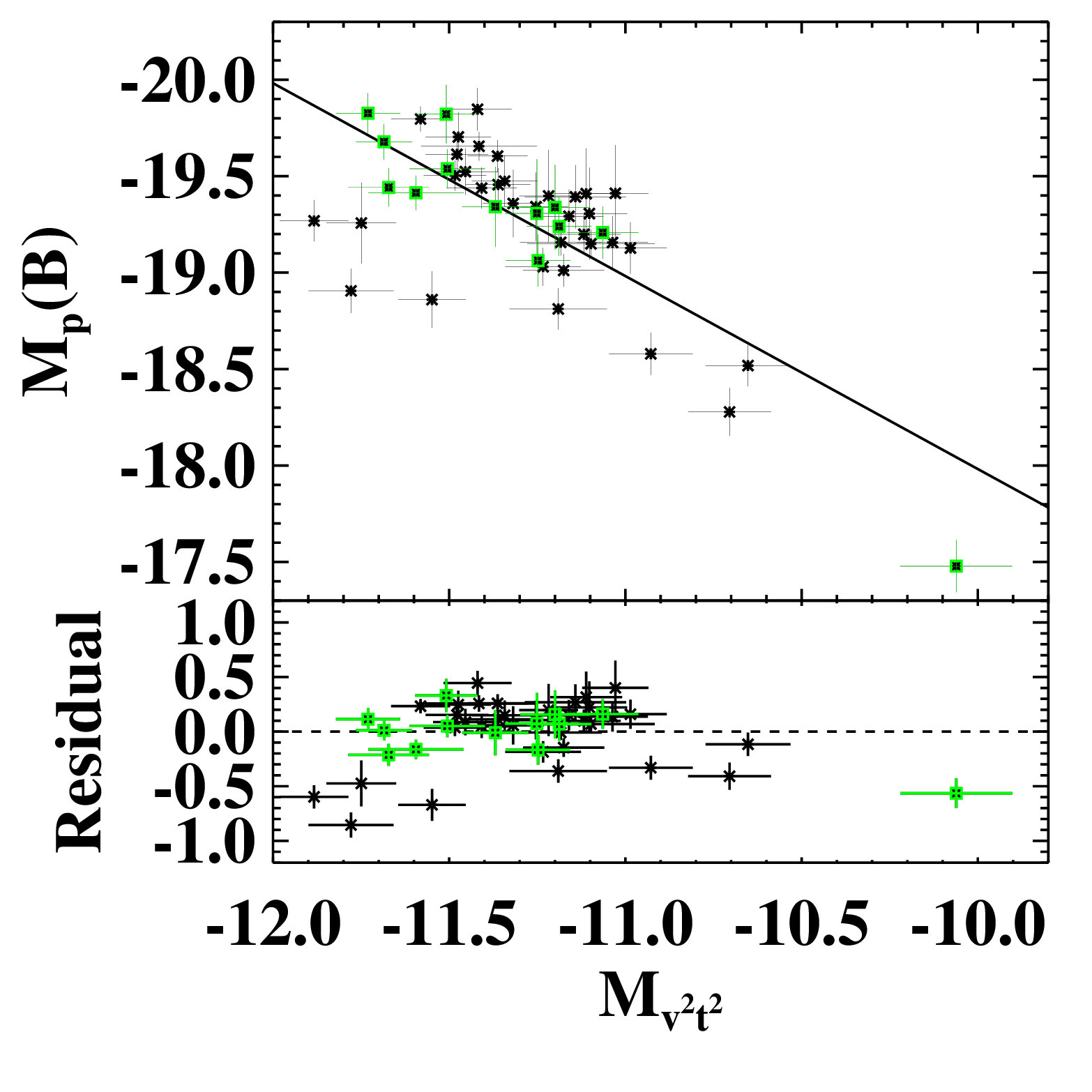}
\includegraphics[width=.320\textwidth]{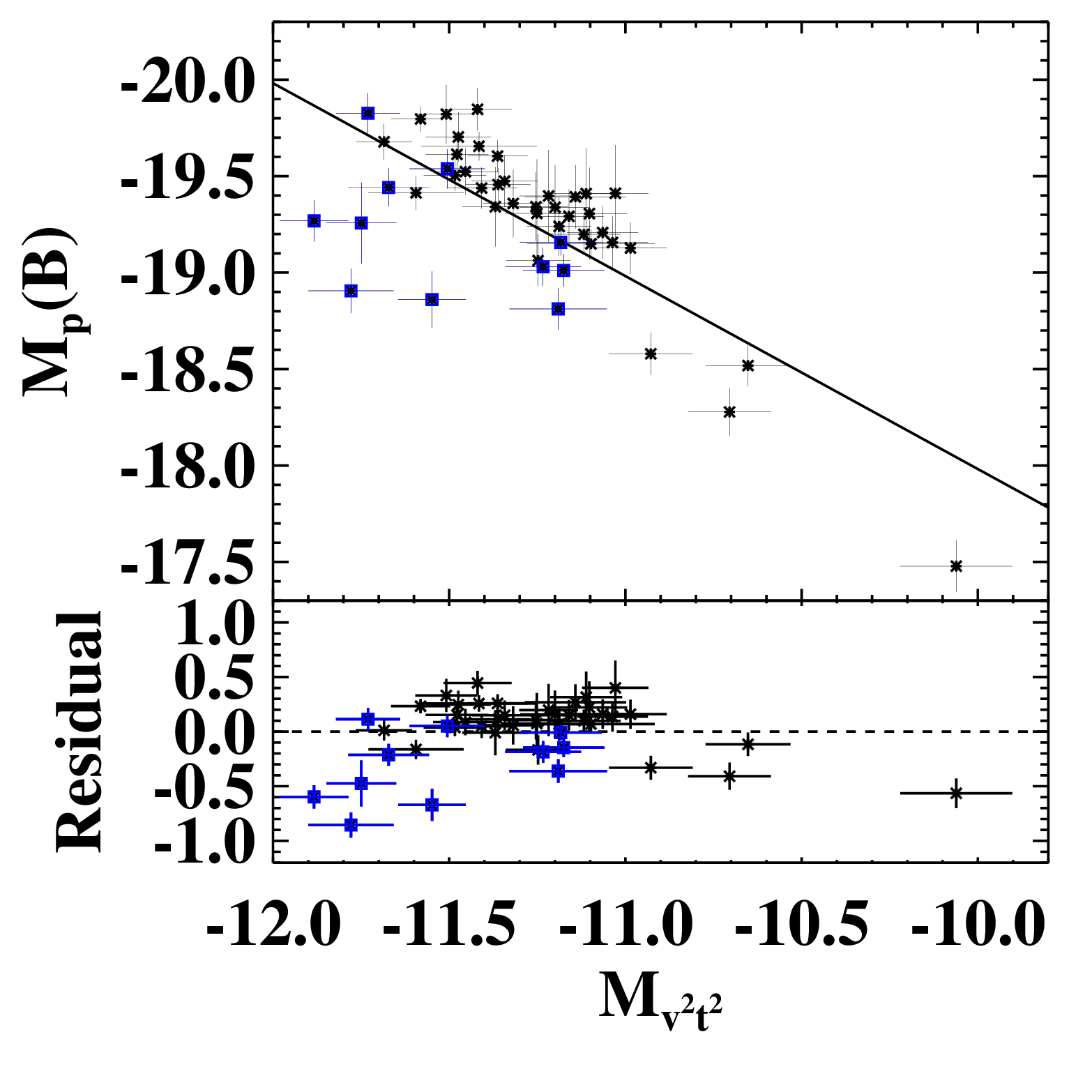}
\caption{Same as the right panel in Figure~\ref{relation_big_sample} for the new 
   $M_p$ vs. $M_{v^2t^2}$
         relation in Equation~\ref{eq_lvt2pm}, 
         but overplotted with different subgroups. 
         Left panel: {\it red} points show the SNe~Ia with large ${\Delta}m_{15}(B)$,
	 which usually do not follow the Phillips relation, but they generally follow
	 the new $M_p$ vs. $M_{v^2t^2}$ relation.
         Middle panel: {\it green} points show the SNe~Ia with medium host-galaxy
	 extinction ($0.1 < E(B-V) < 0.3$~mag).
	 Right panel: {\it blue} points show SNe~Ia having high photospheric 
      velocity,
	 $v_{{\rm Si~II}~\lambda6355} \geq 12.0$k~km~s$^{-1}$ at peak brightness. 
       They appear to be systematically
	 below the best-fit relation.}
\label{relation_bad_sample}
\end{figure*}

First, we consider those SNe~Ia with ${\Delta}m_{15}(B) > 1.6$~mag, which
usually do not follow the Phillips relation well (e.g., Taubenberger 2017).
They are shown as {\it red} points in the left panel of Figure~\ref{relation_bad_sample}.
It is noteworthy that they generally follow the new $M_p$ vs. $M_{v^2t^2}$ 
relation, though they are typically outliers in the Phillips relation.
This means that with the new relation, it is possible to compare them with 
normal SNe~Ia.
However, these SNe are generally below the best-fitting result for the new relation,
indicating that they are likely intrinsically underluminous ---
consistent with the fact that most of them are underluminous SN 1991bg-like 
SNe~Ia.

Next, we examine SNe with different host-galaxy extinctions. Since
the new relation cannot help derive the host extinction, we use the value
estimated from the MLCS2k2 fitting, and we exclude the SNe having large 
extinction ($E(B-V) > 0.3$~mag) in the above studies.
Here we examine those SNe with medium host extinction ($0.1 < E(B-V) < 0.3$~mag), 
which are shown as {\it green} points in the middle panel of Figure~\ref{relation_bad_sample}.
As one can see, these objects spread out in our sample with large scatter, 
consistent with the fact that they have relatively large unknown host-galaxy extinction.

\begin{figure*}
\includegraphics[width=.490\textwidth]{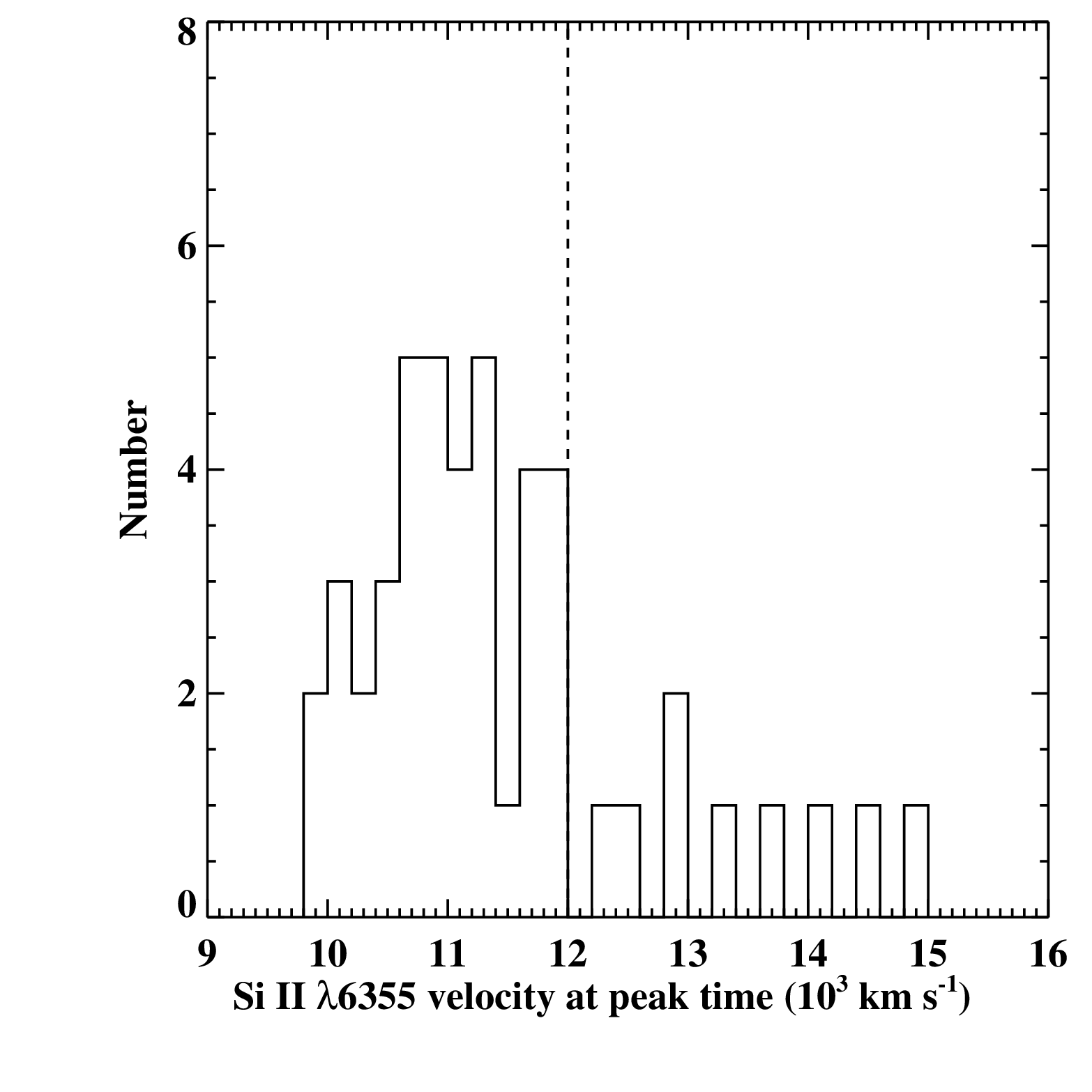}
\includegraphics[width=.490\textwidth]{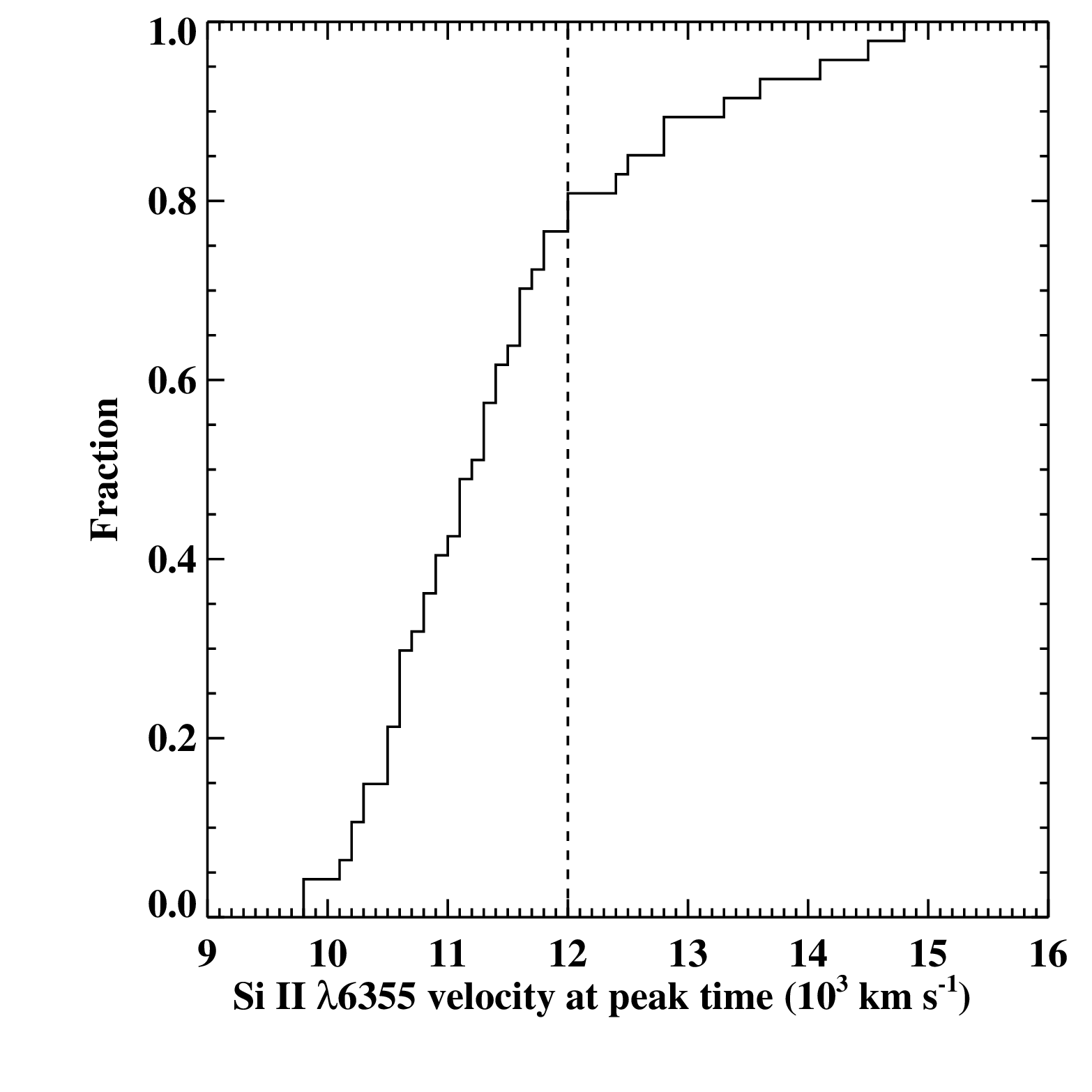}
\caption{Histogram (left panel) and cumulative (right panel) distributions of
         the Si~II $\lambda$6355 velocity at the time of peak brightness 
        for the ``group1'' sample.}
\label{relation_velocity_histgram_cumulative}
\end{figure*}

Lastly, we also examine the SNe with different Si~II $\lambda$6355 velocity at peak brightness.
Wang et al. (2009; 2013) and Foley \& Kasen (2011) grouped high- and 
normal-velocity SNe~Ia based on a photospheric velocity boundary of 
11.8k~km~s$^{-1}$ at peak brightness.
Interestingly, they found that the $B-V$ color at maximum brightness 
of high-velocity SNe~Ia is redder by $\sim 0.1$~mag on average. 
Wang et al. (2013) also found that SNe~Ia with high velocity
($v_{{\rm Si~II}~\lambda6355} \geq 12.0$k~km~s$^{-1}$ at peak brightness)
are substantially more concentrated in the inner and brighter regions of their host galaxies than are
normal-velocity SNe~Ia, and the former tend to inhabit larger and more-luminous hosts. 
Figure~\ref{relation_velocity_histgram_cumulative} shows the histogram and cumulative distributions
of the Si~II $\lambda$6355 velocity at peak brightness for our ``group1'' sample.
Although our sample is not sufficiently large for a double-Gaussian fit like
that of Wang et al. (2013, see their Figure~1$c$),
the high-velocity sample (with $v_{{\rm Si~II}~\lambda6355} \geq 12.0$k~km~s$^{-1}$ 
at peak brightness)\footnote{We could have adopted $v_{{\rm Si~II}~\lambda6355} 
> 11.8$k~km~s$^{-1}$ as done similarly by Wang et al. (2009; 2013)
and Foley \& Kasen (2011), but since no SN in our sample has a velocity
between 11.8k and 12.0k~km~s$^{-1}$, this choice does not affect our sample.},
which constitutes $\sim 1/4$ of our ``group1'' sample,
is clearly distinct from the normal-velocity sample ($v_{{\rm Si~II}~\lambda6355} < 12.0$k~km~s$^{-1}$).
In the right panel of Figure~\ref{relation_bad_sample}, {\it blue} points show the
high-velocity SNe~Ia ($v_{{\rm Si~II}~\lambda6355} \geq 12.0$k~km~s$^{-1}$).
There is a distinct difference between the two groups:
the high-velocity SNe lie systematically below the best-fit $M_p$ vs. $M_{v^2t^2}$ relation,
which means they are probably intrinsically fainter than the normal-velocity SNe~Ia,
if not suffering higher host-galaxy extinction. The higher velocities might
possibly be produced by relatively younger and more metal-rich progenitors 
restricted to galaxies with substantial chemical evolution, as suggested 
by Wang et al. (2013).

It is interesting to see that both subsamples with ${\Delta}m_{15}(B) > 1.6$~mag
(left panel of Figure~\ref{relation_bad_sample}) and with $v_{{\rm Si~II}~\lambda6355} \geq 12.0$k~km~s$^{-1}$
(right panel of Figure~\ref{relation_bad_sample}) are systematically below the best-fit $M_p$ vs. $M_{v^2t^2}$ relation,
probably indicating that they are intrinsically fainter than the normal SNe~Ia.
Our method provides a possible way to distinguish these subluminous SNe~Ia 
from the normal ones.

\subsection{The Low-Velocity Golden Sample}\label{s:goldensample}

\begin{figure*}
\centering
\includegraphics[width=.320\textwidth]{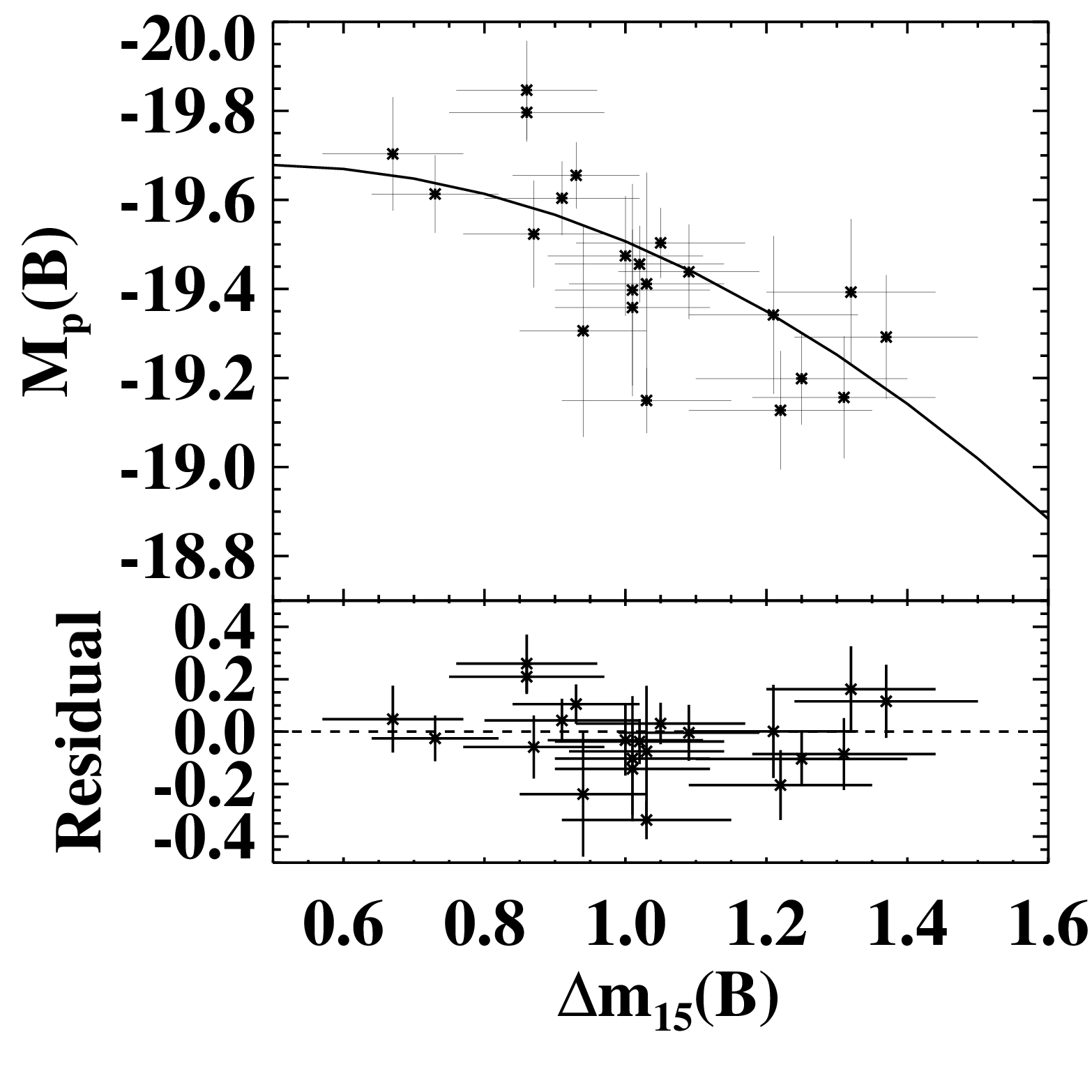}
\includegraphics[width=.320\textwidth]{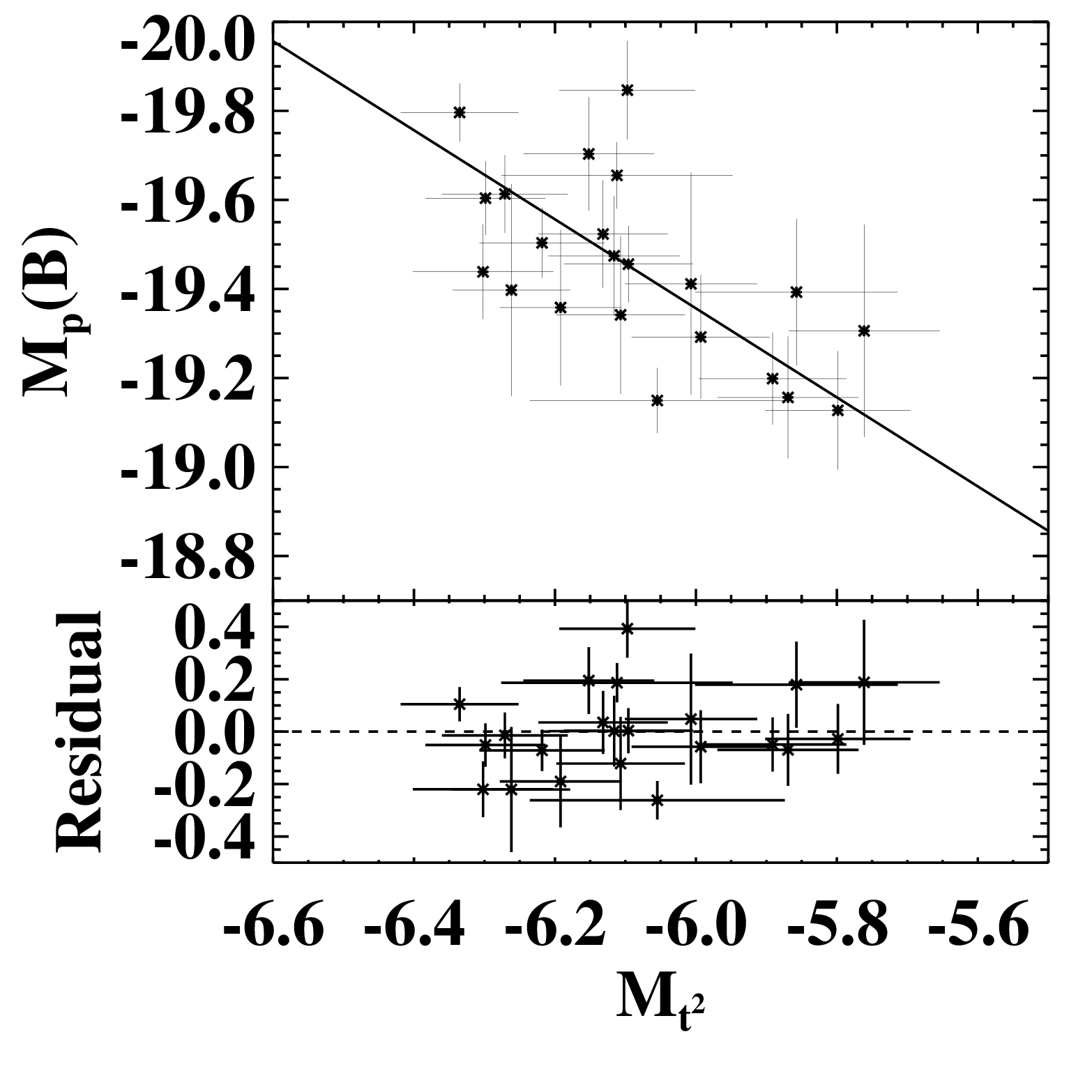}
\includegraphics[width=.320\textwidth]{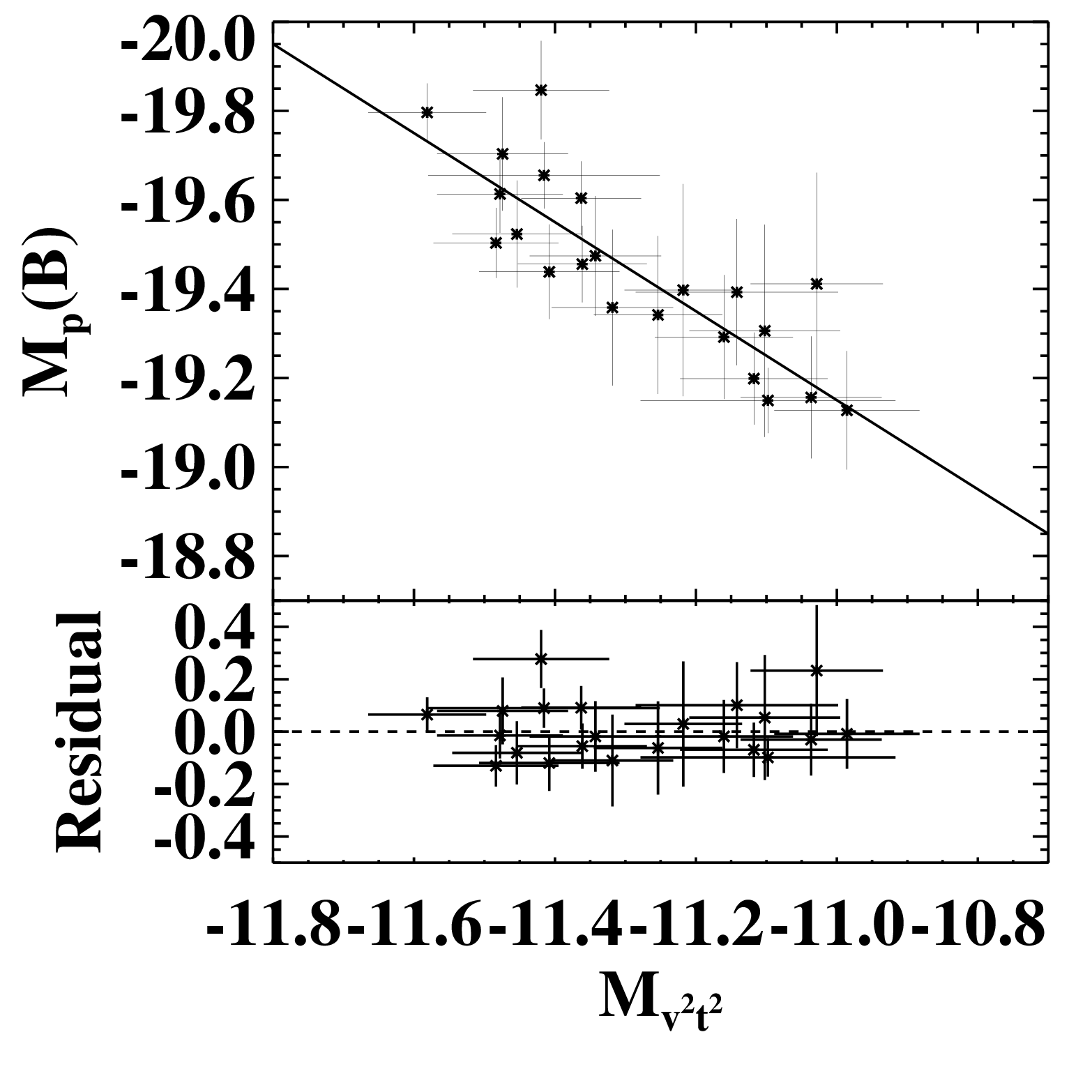}
\caption{Same as Figure~\ref{relation_big_sample}, but for the ``group2'' sample with tighter selection critera;
         see text for details. Compared to Figure~\ref{relation_big_sample}, the scatter
	 in each relation has improved significantly. In particular, the improvement
	 for the new $M_p$ vs. $M_{v^2t^2}$ relation (right panel) is much more 
         significant than that for the others.}
\label{relation_small_sample}
\end{figure*}

In the above section, we show that some subgroups of SNe~Ia are outliers to one or all of the
relations, and some are systematically offset from the best-fit relations. In order to make
a better and more stringent comparison between the different relations, we adopt an
even smaller set of $z > 0.005$ SNe~Ia by excluding those with
${\Delta}m_{15}(B) > 1.6$~mag, those with host extinction $E(B-V) > 0.1$~mag,
and those with $v_{{\rm Si~II}~\lambda6355} \geq 12.0$k~km~s$^{-1}$ at peak brightness.
This set has 22 SNe~Ia (out of the
47 in ``group1''), which we call the ``low-$v$ golden sample'' and label as ``group2'' 
for comparison with further restricted subsets.
This smaller, but more homogeneous, low-$v$ golden sample minimizes the effects of host-galaxy extinction
and other factors (e.g., we exclude outliers to the Phillips relations),
and can therefore better reveal relations between parameters.

\begin{figure*}
\centering
\includegraphics[width=.320\textwidth]{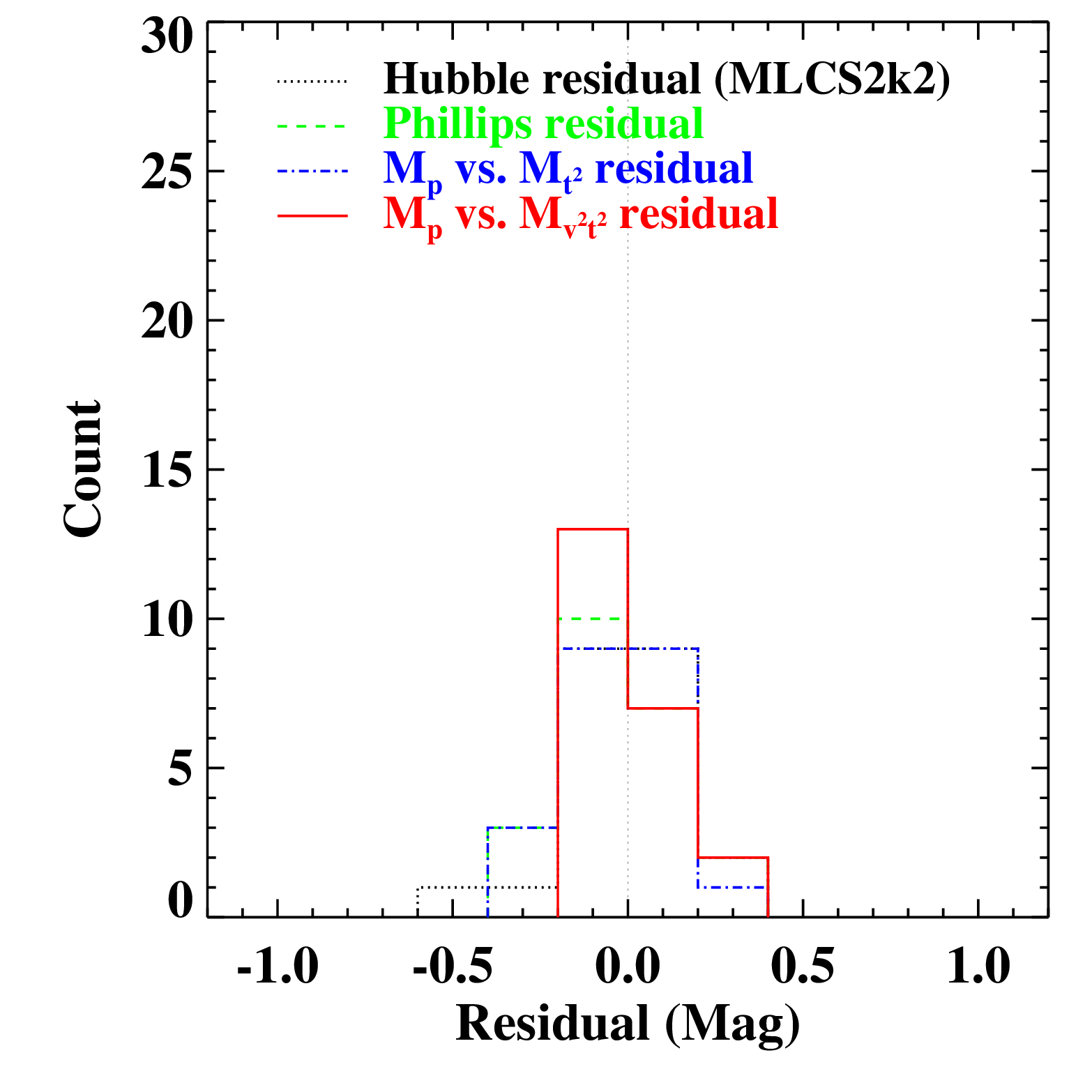}
\includegraphics[width=.320\textwidth]{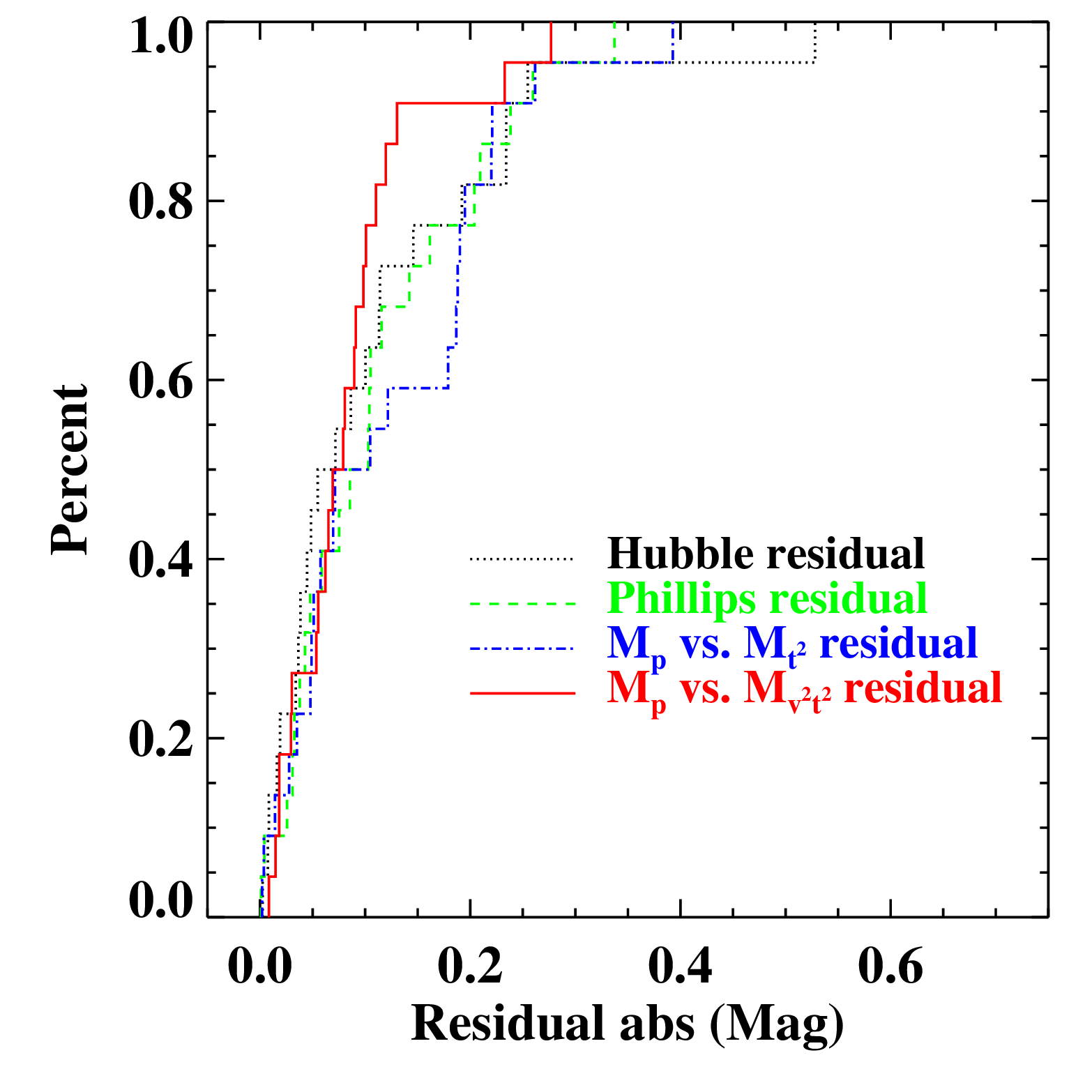}
\includegraphics[width=.320\textwidth]{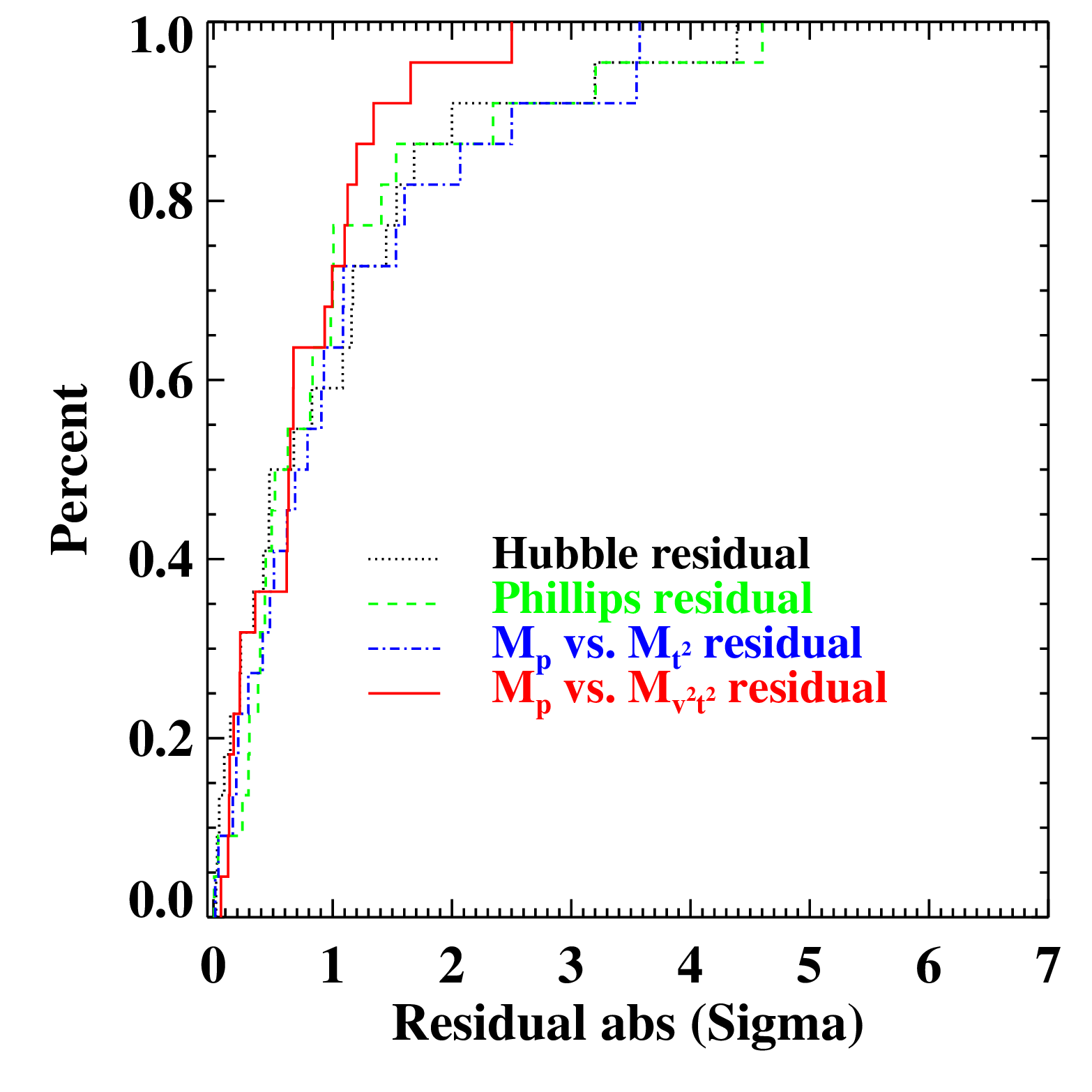}
\caption{Same as Figure~\ref{relation_residual_histgram}, but for ``group2.'' 
     The cumulative distribution clearly shows that the residuals from the new
    $M_p$ vs. $M_{v^2t^2}$ relation (red curve in the middle and right 
    panels) are the smallest.}
\label{relation_residual_histgram_smallsample}
\end{figure*}

\begin{figure*}
\centering
\includegraphics[width=.320\textwidth]{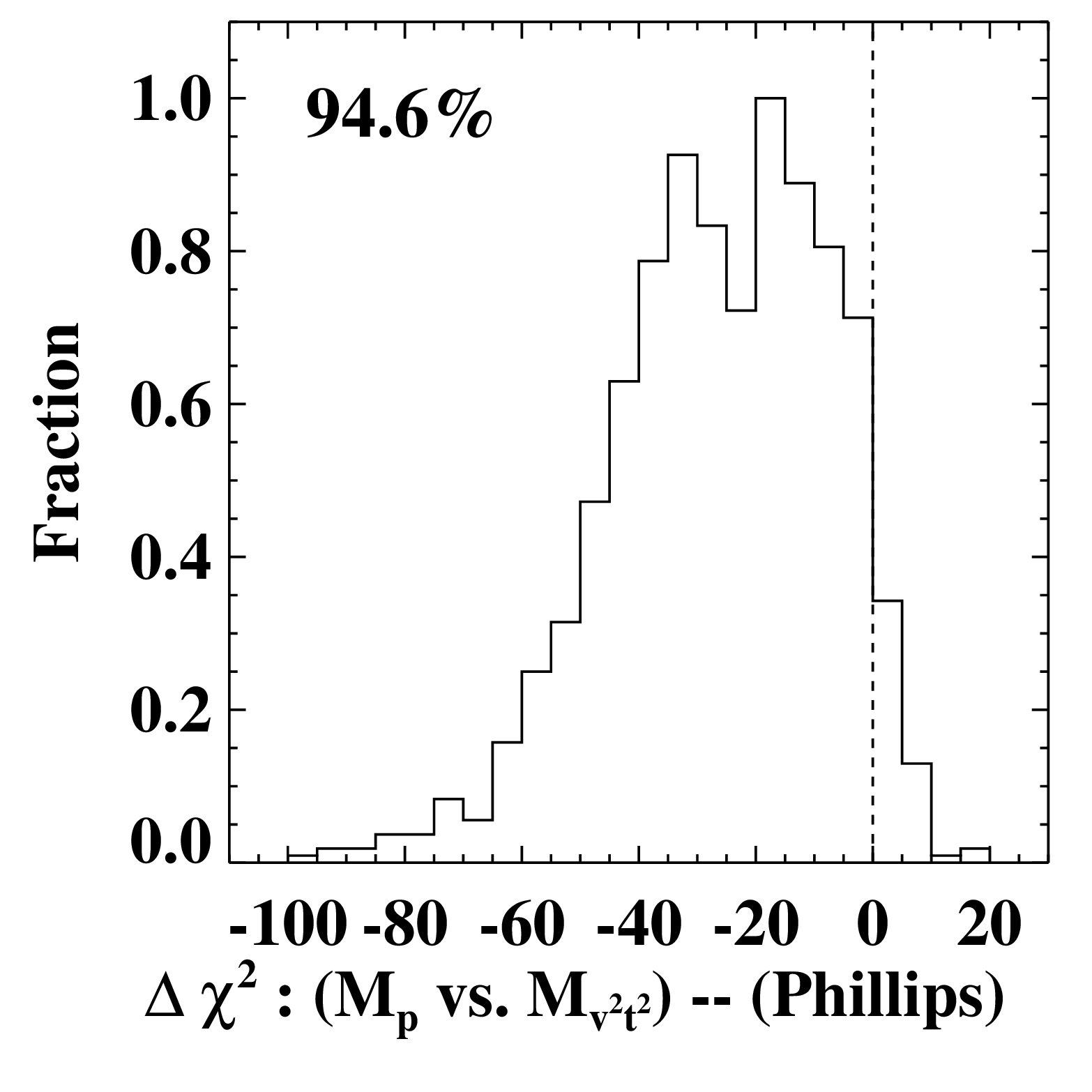}
\includegraphics[width=.320\textwidth]{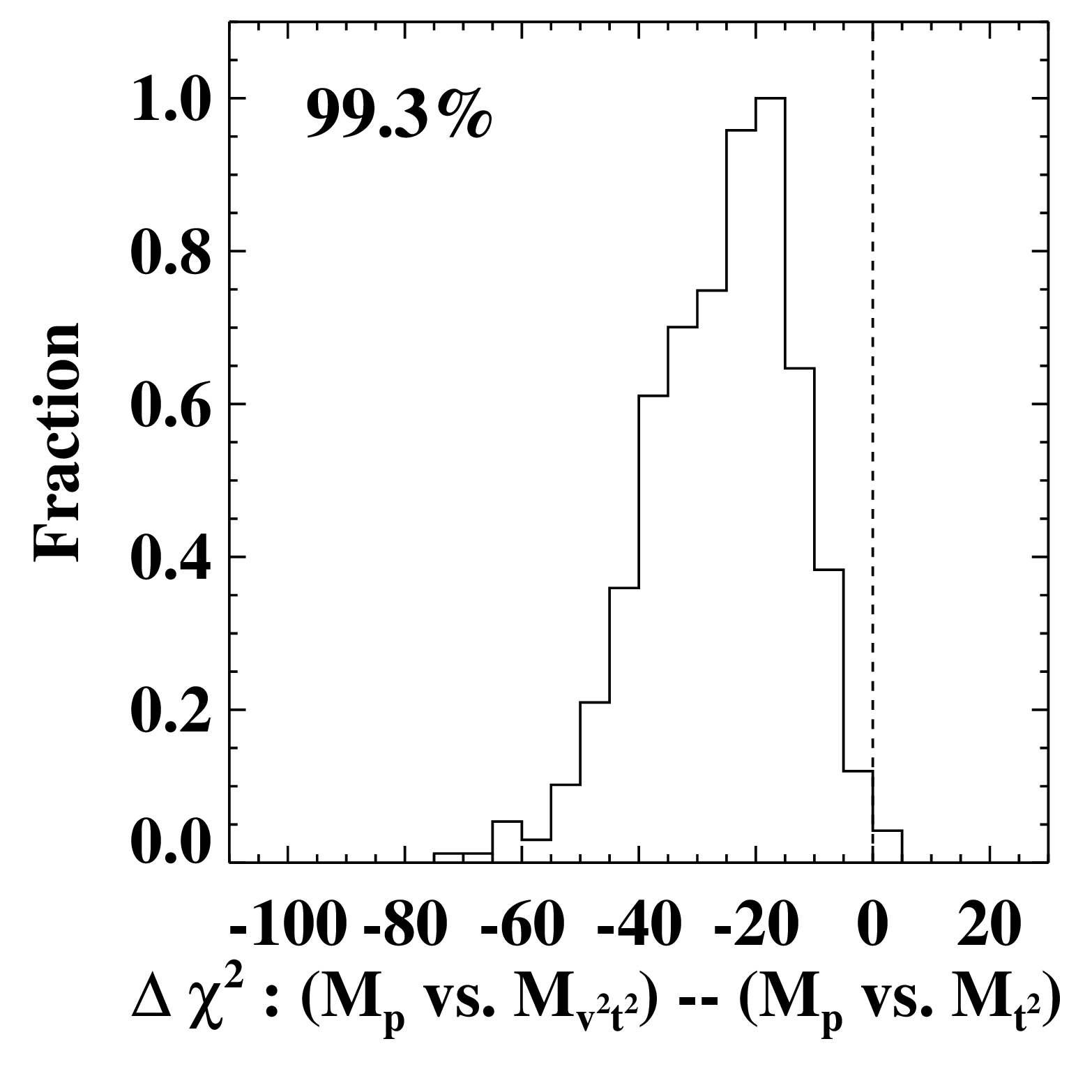}
\includegraphics[width=.320\textwidth]{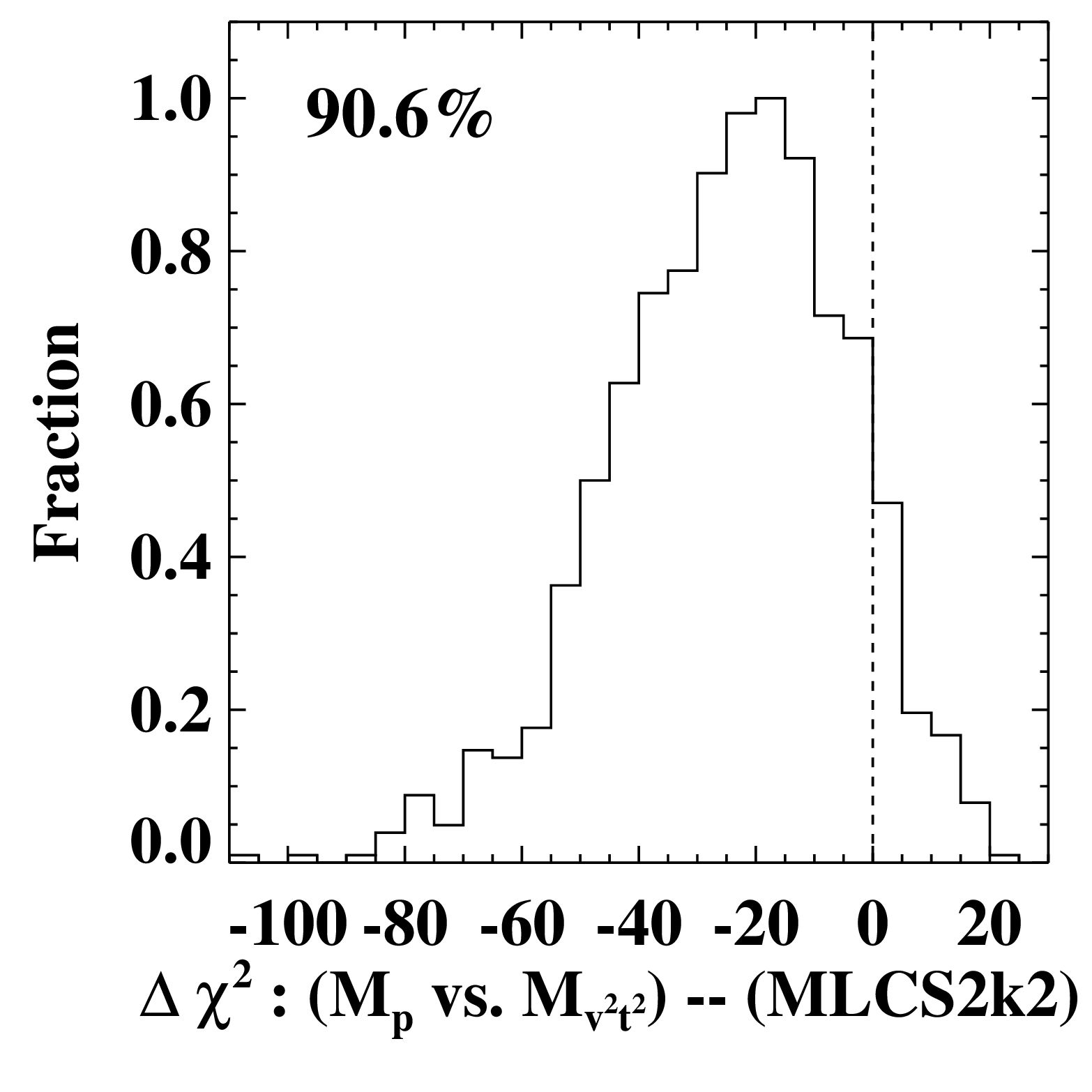}
\caption{Same as Figure~\ref{relation_result_bootstrap_group1}, but for ``group2.''
         The statistical probability shows that the new $M_p$ vs. $M_{v^2t^2}$ 
   relation is much better than all others.\\
         \\
         \\
         \\
         }
\label{relation_result_bootstrap_group2}
\end{figure*}

Similar to ``group1,'' we apply the analysis again to compare the four cases 
(three different relations, as well as the Hubble residuals from MLCS2k2 
fitting) to this low-$v$ golden sample; the results are listed in Table~2
and shown in Figures~\ref{relation_small_sample}--\ref{relation_result_bootstrap_group2}.

The residual scatter and reduced $\chi^2$ of all relations
are improved with the smaller low-$v$ golden sample 
(``group2'') compared with the full sample (``group1''), 
as seen from Figure~\ref{relation_small_sample} and Table~2.
In the case of the ``group1'' sample, the residual scatters are larger than 0.26~mag for
all four cases, while for ``group2,'' the residuals are now all around or below 0.17~mag.
This result is shown in Figure~\ref{relation_residual_histgram_smallsample}, 
where the left panel
shows the histogram distributions of the residuals for all four cases 
and the right panel displays
the corresponding cumulative distributions for ``group2.''
The reduced $\chi^2$ values are also substantially decreased: for ``group1,''
all are higher than 3.4, while now for ``group2,'' all
become smaller than 2.3.

When we use the AIC$_c$ to perform model selection
with the ``group2'' sample, we find very strong positive evidence
in favor of $M_p$ vs. $M_{v^2t^2}$ ($\Delta {\rm AIC}_c \approx 27$), as 
shown in Table~2.
Indeed, while all four relations have similar scatter for ``group1,''  the new 
$M_p$ vs. $M_{v^2t^2}$ relation
yields the smallest scatter among the four cases for ``group2'', 
with a residual of only $0.108 \pm 0.018$~mag,
which is $\sim 0.05$~mag smaller than the other three cases.

This is confirmed by the residual cumulative distributions for the new 
$M_p$ vs. $M_{v^2t^2}$ relation (red curve in the right panel of 
Figure~\ref{relation_residual_histgram_smallsample}), which
clearly stands out at the left edge (i.e., smaller residuals).

We once again use a bootstrap procedure to additionally compare the models as
in Section~\ref{s:relation_full_sample}, but now applying it to ``group2.''
For 1000 simulations, as shown in Figure~\ref{relation_result_bootstrap_group2}, 
the statistical probability is 94.6\%, 99.3\%, and 90.6\% better than the Phillips relation,
the $M_p$ vs. $M_{t^2}$ relation, and the MLCS2k2 fitting method, respectively.

Although we have removed peculiar velocities expected from 
the Carrick et al. (2015) model, we expect that the model is 
imperfect. There should be significant residual scatter arising
from peculiar velocities of $\sim250$\,km s$^{-1}$.
We note that we have not removed this contribution from 
the residual scatter we have presented in this paper.
Our residual from the new $M_p$ vs. $M_{v^2t^2}$ relation for ``group2'' 
is only $0.108 \pm 0.018$~mag. If we adopt a median residual peculiar-velocity 
uncertainty of $\pm 250$~km~s$^{-1}$, corresponding to 0.11~mag, we
conclude that the measured residual scatter for the new $M_p$ vs. $M_{v^2t^2}$ 
relation is likely dominated by the peculiar-velocity uncertainty.
A more precise Monte Carlo simulation, as in Section~\ref{s:relation_full_sample}, 
gives a scatter of $0.135 \pm 0.030$~mag with an average 
peculiar velocity of $\pm 250$~km~s$^{-1}$ applied to this low-$v$ golden sample.
We note that we obtain a similar intrinsic scatter if we do not
remove peculiar velocities expected from the Carrick et al. (2015) 
model, and assume a scatter arising from peculiar motions of
300\,km\,s$^{-1}$.

Another factor that may contribute to the final scatter is the 
temperature of SNe~Ia at the time of peak brightness.
In Equation~\ref{eq_lvt2p}, we assume that SNe~Ia all have the same 
temperature at peak brightness, but this is not actually true.
If the temperature among SNe~Ia varies by 10\%, 
this will contribute $\sim 0.1$~mag to the final scatter, 
assuming optical passbands lie on the Rayleigh-Jeans tail of SN spectra.
For extreme cases, comparing the SN 1991bg-like SNe~Ia with SN 1991T-like SNe~Ia, 
adopting temperatures from Nugent et al. (1995) could cause a 0.43~mag difference.
On average, SN 1991bg-like SNe~Ia are $\sim 0.22$ mag fainter than normal SNe~Ia,
which could partially explain why SN 1991bg-like SNe~Ia are generally below the
best-fit line in Figure~\ref{relation_bad_sample} (left panel), but 
cannot explain the high-velocity SNe~Ia that are also below the
best-fit line in Figure~\ref{relation_bad_sample} (right panel).
Note that previous works have found (e.g., Wang et al. 2009; Foley et al. 2011;
Blondin et al. 2012; Mandel et al. 2014) that high-velocity SNe~Ia are
likely intrinsically redder (lower temperature) than normal-velocity SNe~Ia,
but the difference is too small to account for the $\sim 0.4$~mag we see
in the residual in Figure~\ref{relation_bad_sample} (right panel).
On the other hand, as discussed earlier, the $B$ filter has a central wavelength
that is too blue to fall on the Rayleigh-Jeans tail, so the SN~Ia luminosity would increase
much more rapidly than linearly with the temperature as we assumed; in that case, the 
difference caused by the temperature would become much larger than indicated above.
Therefore, we cannot completely rule out the possibility that temperature differences
are responsible for the offsets of high-velocity SNe~Ia.
Moreover, a blackbody is not expected to provide an accurate model for the  
SN emission near maximum light.
Since the scatter caused by the temperature is included in the final
dispersion, it is difficult to distinguish this component without detailed
modeling to derive the temperature, which is beyond the scope of
this paper.

We note that the MLCS2k2 estimate of $A_V$ may also 
compensate for color variation intrinsic to the 
SN, that may result from temperature variation.
We attempt to estimate the scatter from the temperature
contribution by searching for correlations between the residual 
and the color at peak brightness (e.g., $B-V$), but we
do not see any clear trend in our sample.
It is possible that independent constraints on the effective
temperature of the SN photosphere could be helpful
in improving the calibration, and to disentangling the effect
of intrinsic color variation and dust extinction (see, e.g., Scolnic et al. 2014; Mandel et al. 2016).

\subsection{Subgroups in the Low-Velocity Golden Sample}

\begin{figure}
\centering
\includegraphics[width=.490\textwidth]{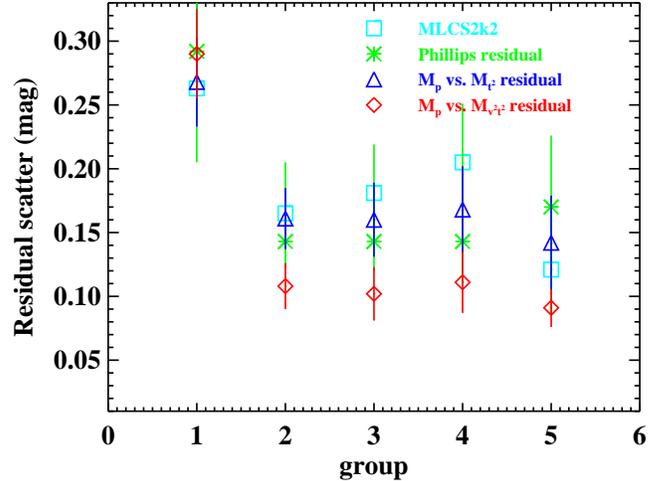}
\caption{The residual scatter for the low-$v$ golden sample changes 
       with redshift cutoff from ``group2'' ($z > 0.005$) through
       ``group5'' ($z > 0.020$). Also shown is the result
         for the full sample of ``group1.''}
\label{relation_residual_z_changes}
\end{figure}

As discussed above, the dispersion in inferred SN distances arising from the residual peculiar velocities likely
contributes substantially to the final scatter. The scatter contributed 
by peculiar velocities, however, should decrease with increasing redshift. To 
assess this expectation,
we divide the low-$v$ golden sample into a few subgroups. 
For ``group3,'' we select only SNe with $z > 0.010$ from ``group2''
($z > 0.005$).
Similarly, we select a ``group4'' sample with $z > 0.015$ and a 
``group5'' sample with $z > 0.020$.
We applied the same analysis as for ``group2'' and list the results in Table~2.
They confirm that the scatter from the peculiar velocities decreases 
from ``group2'' to ``group5'' as the sample
redshift increases (Figure~\ref{relation_residual_z_changes}). 
In general, the scatter for all of the relations 
decreases as the sample redshift increases; however,
the new $M_p$ vs. $M_{v^2t^2}$ relation gives the
best result (smallest scatter) among the four cases.
This is clearly shown in Table 2 by the AIC$_c$ statistic comparison
(for our cases, since the number of parameters is the same, comparing AIC$_c$ is equivalent to
comparing the $\chi^2$ value).
For ``group3,'' ``group4,'' and ``group5,'' the difference between the AIC$_c$ value for $M_p$ vs. $M_{v^2t^2}$ and the other relations is at least 6, which indicates significant positive evidence.
The bootstrap simulations yield a consistent, if less significant, result: the $M_p$ vs. $M_{v^2t^2}$ relation yields smaller scatter for 93.3\%, 78.7\%, and 79.3\% of bootstrap realizations than the Phillips relation
for ``group3,'' ``group4,'' and ``group5,'' respectively.
As compared with the MLCS2k2 method, the probability is
89.5\%, 85.8\%, and 68.1\% better (respectively); however, we note that the sample sizes for group4 and group5 are relatively small.

\section{Discussion}\label{s:discussion}

\begin{deluxetable}{ccc|cc}
 \tabcolsep 0.4mm
 \tablewidth{0pt}
 \tablecaption{Fitting Comparison}
  \tablehead{\colhead{Groups} & \colhead{$\chi^2$}       & \colhead{AIC$_c$} & \colhead{$\chi^2$} & \colhead{AIC$_c$}}
\startdata
        & \multicolumn{4}{c}{Phillips relation} \\
\hline
        & \multicolumn{2}{c}{$k=3$} & \multicolumn{2}{c}{$k=1$} \\
group1  &  217.23 &  223.79   &  240.00  &   242.09    \\
group2  &  41.49  &  48.52    &  47.42   &   49.62     \\
group3  &  40.36  &  47.96    &  46.11   &   48.35     \\
group4  &  36.79  &  45.19    &  41.49   &   43.82     \\
group5  &  18.52  &  32.52    &  34.08   &   36.88     \\
\\
\hline
        & \multicolumn{4}{c}{$M_p$ vs. $M_{v^2t^2}$ relation} \\
\hline
        & \multicolumn{2}{c}{$k=2$} & \multicolumn{2}{c}{$k=1$} \\
group1  &  262.26 &  266.53   &   263.27 &   265.36    \\
group2  &  18.34  &  22.97    &   19.35  &   21.55     \\
group3  &  16.86  &  21.61    &   18.40  &   20.64     \\
group4  &  15.34  &  20.43    &   17.46  &   19.79     \\
group5  &  7.08   &  14.08    &   8.61   &   11.41     \\
\enddata
\end{deluxetable}

Our results are robust to the specific choice of sample extinction, redshift, and velocity cuts that we adopt. 
Wang et al. (2009, 2013) and Foley \& Kasen (2011) apply similar photospheric
velocity cuts to study populations of low- and high-velocity SNe.

We note that there is an outlier, SN~2000dn, quite far away from the Phillips relation.
This SN is more consistent with the new $M_p$ vs. $M_{v^2t^2}$ relation.
To test whether our results were robust to excluding this SN, we refit all the group samples after removing it.
Table 4 shows the $\chi^2$ results for comparison.
The $M_p$ vs. $M_{v^2t^2}$ relation still offers significant improvement when compared to the Phillips relation for ``group2'' and ``group3,'' although we find no difference for the smaller ``group4'' and ``group5'' samples.

\begin{deluxetable}{ccccc}
 \tabcolsep 0.4mm
 \tablewidth{0pt}
 \tablecaption{$\chi^2$ Fitting Results After Excluding SN~2000dn}
  \tablehead{\colhead{Groups} & \colhead{Hubble residual             } & \colhead{Phillips         } & \colhead{$M_p$ vs. $M_{t^2}$         } & \colhead{$M_p$ vs. $M_{v^2t^2}$         }}
\startdata
group1  &  148.15  &  225.96  &  232.39  &  262.32  \\
group2  &  35.34   &   23.97  &   33.04  &   17.35  \\
group3  &  35.25   &   22.06  &   31.49  &   16.51  \\
group4  &  33.01   &   15.76  &   27.54  &   15.46  \\
group5  &  4.24    &   7.90   &   8.61   &    6.62 
\enddata
\end{deluxetable}


\section{Conclusions}\label{s:conclusions}

From the above analysis for both the ``group1'' and ``group2'' samples, 
we obtain the following conclusions.

(1)
The $M_p$ vs. $M_{v^2t^2}$ relation yields the most precise distances among the four models considered for the low-$v$ golden sample.

(2) The new $M_p$ vs. $M_{v^2t^2}$ relation is mathematically 
linear, as shown in Equation~\ref{eq_lvt2pm}; this
offers a useful simplification compared to the quadratic Phillips relation.

(3) The rise time $t_r$ is probably better than the decay-time 
parameter ${\Delta}m_{15}(B)$ for studying SNe~Ia. Both the
$M_p$ vs. $M_{v^2t^2}$ and the $M_p$ vs. $M_{t^2}$  relations are 
comparable to (or much better than) the Phillips relation; 
moreover, they are easier to explain with linear formulas 
(Equations~\ref{eq_lvt2pm} and \ref{eq_lt2pm}).

(4) The photospheric velocity plays an important role in improving 
estimates of SN~Ia luminosities within our model.
Comparing the $M_p$ vs. $M_{t^2}$ relation (without 
considering the photospheric velocity) with the new $M_p$ vs. $M_{v^2t^2}$ 
relation (considering the photospheric velocity), we see that the
latter is significantly better than the former, especially for the
``group2'' samples.

(5) We also confirm that high-velocity SNe~Ia are probably
intrinsically different from normal-velocity SNe~Ia, consistent with
the conclusions of other groups (Wang et al. 2009, 2013; Foley \& Kasen 2011).
We show that high-velocity SNe~Ia are probably intrinsically
fainter than the normal-velocity SNe~Ia.

In the future, the new $M_p$ vs. $M_{v^2t^2}$ relation can 
potentially be extended to higher redshifts. If the SN light curves
are well observed, then the first-light times can be estimated
with the method presented by Zheng \& Filippenko (2017) and Zheng at al. (2017). This may lead 
to another way of determining SN distances and be used for cosmology 
(e.g., Jha et al. 2007; Guy et al. 2007).
Since our method relies on the estimate of first-light time, it is
very important to obtain a good light-curve sample in the $B$ (or $g$) band.
Our suggestion for future surveys is to perform
high-cadence photometry (ideally daily, if possible, or at least every
other day), and to obtain at least one spectrum near maximum light.

To conclude, we have examined a new three-parameter relationship in 
Type Ia SNe: peak magnitude, rise time, and photospheric 
velocity. This $M_p$ vs. $M_{v^2t^2}$ relation is based on observations,
though it is motivated by (and physically easy to explain with) the simple fireball model.
We compared it with other SN~Ia relations and found smaller scatter;
thus, it has the potential to be used for accurate cosmological distance
determinations.

\begin{acknowledgments}

We thank Isaac Shivvers, Melissa L. Graham, and an anonymous referee for useful discussions
and suggestions. A.V.F.'s supernova group at UC Berkeley is grateful for financial
assistance from NSF grant AST-1211916, the TABASGO Foundation,
the Christopher R. Redlich Fund, and the Miller Institute for Basic
Research in Science (U.C. Berkeley).
Research at Lick Observatory is partially supported by a generous gift from Google.

\end{acknowledgments}

\end{document}